\documentclass[12pt]{article}

\usepackage{amsmath, amssymb, amsfonts}
\usepackage{graphicx}
\usepackage{geometry}
\usepackage{setspace}
\usepackage{amsthm}  

\theoremstyle{plain}
\newtheorem{theorem}{Theorem}

\newtheorem{corollary}[theorem]{Corollary}
\newtheorem{remark}[theorem]{Remark}

\usepackage{enumitem}

\newtheorem{lemma}{Lemma}

\usepackage{caption}

\usepackage{booktabs}        
\usepackage{caption}         

\title{\textbf {A General Theory of Growth, Employment, and Technological Change: Experiential Matrix Theory and the Transition from GDP to Humanist Experiential Growth in the Age of Artificial Intelligence}}
\author{Christian William Callaghan\\Working Paper 2025-1-20\\Economics of Artificial Intelligence Research Laboratory\\Centre for Inclusive Societies and Economies\\Anglia Ruskin University\\Cambridge\\United Kingdom}
\date{\relax}

\begin{document}

\maketitle

\begin{abstract}
This paper introduces \textit{Experiential Matrix Theory} (EMT), a general theory of growth, employment, and technological change for the age of artificial intelligence (AI). EMT redefines utility as the alignment between production and an evolving, infinite-dimensional matrix of human experiential needs, thereby extending classical utility frameworks and integrating ideas of the capabilities approach of Sen and Nussbaum into formal economic optimisation modelling. 

We model the economy as a dynamic control system in which AI collapses ideation and coordination costs, transforming production into a real-time vector of experience-aligned outputs. Under this structure, the production function becomes a continuously learning map from goods to experiential utility, and economic success is redefined as convergence toward an asymptotic utility frontier in $\ell^\infty$ space. Using Pontryagin’s Maximum Principle in an infinite-dimensional setting, we derive conditions under which AI-aligned output paths are asymptotically optimal, and prove that unemployment is Pareto-inefficient wherever unmet needs and idle human capacities persist.

On this foundation, we establish \textit{Alignment Economics} as a new research field dedicated to understanding and designing economic systems in which technological, institutional, and ethical architectures co-evolve. EMT thereby reframes policy, welfare, and coordination as problems of dynamic alignment, not static allocation, and provides a mathematically defensible framework for realigning economic production with human flourishing. As ideation costs collapse and new experiential needs become addressable, EMT shows that economic growth can evolve into an inclusive, meaning-centred process—formally grounded, ethically structured, and AI-enabled.\footnote{This paper presents the mathematical formalisation of Experiential Matrix Theory (EMT). A complementary version focusing on the theoretical foundations and social science framing is available on SSRN.}\footnote{Disclaimer: Mathematically derived claims depend on assumptions. Mathematics in this paper is employed in a stylised and heuristic manner, with all formal claims derived from the foundational assumptions of EMT. It is acknowledged that these foundational assumptions are not themselves modelled endogenously within the paper, and that they rest upon deeper philosophical and normative commitments which are also not modelled. This creates a structured hierarchy of assumptions. For the purposes of delimitation, several of EMT’s core assumptions are taken as necessary and exogenous to the modelling effort. These include, but are not limited to: (i) that human experiential needs are effectively unlimited and continuously evolving; (ii) that both physical and virtual production systems can be aligned with these needs; (iii) that AI reduces the marginal cost of ideation and production asymptotically; and (iv) that full human employment is a structurally stabilising feature of complex economic systems. 

This paper is one of a coordinated set of preprints forming part of a larger theoretical programme aimed at reforming understanding of how AI can contribute to growth, employment, and responsible and ethical technological change. All models are stylised, and subject to refinement through scholarly discourse and empirical follow-up. Accordingly, all mathematical derivations and formal conclusions presented here hold conditionally, i.e., they are valid only to the extent that EMT’s theoretical framework remains conceptually coherent and its assumptions are accepted. The purpose of the modelling is thus not to offer empirical verification, but to test whether EMT’s normative commitments can be internally reconciled within the formal constraints of neoclassical economic logic.
Consequently, all modelling is based on stylised and idealised assumptions and is intended to function heuristically, to support conceptual exploration rather than to solve real-world problems or render them tractable. The mathematics should be interpreted as a formal scaffolding for theory-building, not as empirically calibrated or predictive modelling. EMT is therefore offered as a generative structure to provoke new directions in economic thought. No empirical claims are made, and all conclusions are conditional upon the theoretical structure defined herein. Claims are thus understood as predictions, including policy recommendations. Empirical testing and validation of theoretical ideas is left for future work. 
}\footnote{Disclaimer- Acknowledging the humanist normative framing of EMT: This paper adopts a normative, humanist framing that departs from the strictly non-normative stance typically associated with much of mathematical economics. Specifically, EMT is grounded in the proposition that economic systems should centre on the alignment of production with evolving human experiential needs. As such, EMT explores the theoretical plausibility of reconciling full human employment with full AI-enabled production. 

This orientation introduces a potential normative bias, namely, that unemployment, in the presence of idle labour and unmet experiential needs, is both structurally irrational and Pareto inefficient. However, rather than allowing this bias to dominate the analysis, EMT deliberately adopts formal mathematical modelling as a corrective discipline. By embedding its arguments within the logic of neoclassical utility optimisation and general equilibrium frameworks, the theory rigorously tests whether its human-centred outcomes can be derived without abandoning the canonical structure of economic reasoning.

Thus, EMT does not reject the formal structure of mainstream economics, but uses it as a heuristic and constraint, seeking to extend its applicability, not disrupt it. The result is a humanist theoretical architecture that remains grounded in the price-theoretic foundations of economic modelling, while probing whether such models can be reoriented to serve a broader conception of welfare and employment in a context of evolving AI capabilities.}



\end{abstract}

\vspace{1em}
\vspace{1em}
\noindent\textit{Keywords:} economic growth; technological change; artificial intelligence; infinite-dimensional utility; experiential needs; optimal control; endogenous adaptation; post-scarcity.

\noindent\textit{JEL classification codes:} O41; E24; D83; O33; C61; O10.

\section{Introduction}

Standard economic theory frames the objective of growth as the expansion of material output, governed by constraints of resource scarcity and guided by preferences encoded in static utility functions. Yet, as artificial intelligence (AI) radically reduces the cost of ideation, prediction, and coordination, the foundational assumptions underpinning this framework are destabilised. The production function itself becomes endogenous, adaptive, and increasingly decoupled from the classical bottlenecks that once defined economic possibility. This paper introduces Experiential Matrix Theory (EMT), a general theory of growth, employment, and technological change that redefines economic success as the alignment of production with an evolving, infinite-dimensional matrix of human needs.

The contribution of EMT is to formalise utility not as a fixed preference ordering over goods, but as a dynamic mapping between output and lived human experience. Drawing on and extending the utility frameworks of von Neumann and Morgenstern (1944) and Debreu (1959), EMT embeds experiential, psychological, and existential needs directly into the utility function. It integrates capability theory (Sen, 1996; Nussbaum, 2007) with optimal control theory to offer a mathematical structure in which AI-enabled economies progressively learn to align production with utility-bearing dimensions that are typically excluded from market signalling, such as belonging, purpose, or ecological sustainability.

In doing so, the paper addresses a series of unresolved theoretical tensions. First, it explains how long-standing disjunctions between production and well-being that are central to critiques of gross domestic product (GDP) and revealed preference models can be resolved through AI-mediated feedback mechanisms. 

Second, it confronts the paradox of unemployment in an economy with unfulfilled human needs. By showing that unemployment is Pareto-inefficient wherever AI-enabled systems can identify, model, and meet those needs, EMT reframes full employment as a structural requirement for maintaining experiential alignment rather than a Keynesian policy lever.

Third, EMT extends endogenous growth theory (Romer, 1990; Jones, 1995) by proposing that AI collapses the cost not only of production but of ideation itself, thus expanding the feasible set of satisfiable needs in real time. The economy is modelled as a learning system, in which the transformation function from goods to experiences becomes increasingly efficient, eventually converging to the full experiential matrix. This convergence is formally demonstrated using Pontryagin’s Maximum Principle in infinite-dimensional space, yielding necessary conditions under which AI-aligned production maximises experiential utility over time.

Finally, the paper introduces the concept of Alignment Economics, a new research programme grounded within the Samuelson (1947) framework of economics, where coordination occurs through decentralised price signals within functioning market systems. Rather than abandoning the core apparatus of general equilibrium or price-based allocation, EMT builds upon it, extending its scope to incorporate dynamic feedback from ethical, institutional, and technological domains. It formalises how AI, by collapsing ideation costs and enhancing information flow, can augment the market’s capacity to align output with evolving human needs. In this setting, AI does not replace the market, but enhances its responsiveness and adaptivity, functioning as a decentralised planner embedded within the price system. EMT thus provides the theoretical foundation for evaluating AI’s role not only as a productivity enhancer but as a coordination mechanism capable of aligning production with the full structure of human well-being. In this view, meaning, dignity, and existential continuity are not treated as externalities but as endogenous, utility-bearing components central to the definition of economic value.

By placing AI at the heart of the production-to-utility transformation, EMT offers a structural framework for post-scarcity economic design, one in which the economy's objective is not merely to allocate goods efficiently but to expand the frontier of satisfiable human experience. It reframes the purpose of economics as the ongoing alignment of intelligent systems with evolving human needs, thereby redefining what it means for an economy to grow.

\subsection{Contributions}

This paper contributes to economic theory by introducing EMT, a framework that redefines utility as an evolving, infinite-dimensional construct, thereby extending and integrating several foundational strands of economic thought. The work, therefore, seeks to make the following contributions. 

First, it extends ideas from foundational utility theory developed by von Neumann and Morgenstern (1944) and the axiomatic representation of preferences articulated by Debreu (1954, 1959) by situating utility not only in terms of individual choice under uncertainty but within an evolving infinite-dimensional experiential matrix of human needs. EMT also draws on the human capability frameworks of Sen and Nussbaum, integrating their emphasis on dignity, freedom, and functionings into a formal economic optimisation logic. Crucially, EMT argues that traditional measures such as GDP, while capturing physical output, may reflect declining productivity growth over time even as the experiential utility frontier expands exponentially. This disconnect arises because GDP does not account for non-market experiential goods, such as meaning, identity, emotional well-being, or creativity, which increasingly define human welfare in AI-augmented economies. As a result, EMT repositions full employment not as a Keynesian corrective or political ideal, but as a structural necessity for sustaining alignment between AI-enabled production and the evolving complexity of human utility.

Second, EMT reinterprets the role of technology in economic growth. While Romer's (1990) endogenous growth model emphasises ideas as drivers of growth, EMT introduces the concept of AI-induced ideation cost collapse, where the cost of generating and implementing new ideas declines exponentially over time. This notion extends the general-purpose technology framework of Bresnahan and Trajtenberg (1995) by positioning AI not just as a productivity enhancer, but as a catalyst for aligning production with the full spectrum of human experiential needs.

Third, EMT transforms the traditional production function by mapping production vectors to experiential states through an AI-augmented transformation. This approach seeks to generalise Becker's (1976) insights on non-material utility and incorporates real-time sensing and adaptive responsiveness to extend work such as Acemoglu and Restrepo's (2019) analysis of AI's economic impacts. 

Fourth, EMT contributes to the field of optimal control theory by applying Pontryagin's Maximum Principle (Pontryagin et al., 1962) and the infinite-horizon control framework of Balder (1983) to model convergence of AI-aligned production to an ideal experiential matrix. This application extends ideas from the Ramsey-Cass-Koopmans model (Ramsey, 1928; Cass, 1965; Koopmans, 1965) into the domain of evolving experiential utility.

Finally, EMT introduces the concept of alignment economics, a new field that integrates AI, institutional design, and economic production systems. This field generalises ideas drawn from the endogenous growth frameworks of Jones (1995) and Arthur's (1989) increasing returns model, incorporating ethical considerations and the co-evolution of technology and human values.

Through these extensions, EMT provides a comprehensive and mathematically stylised framework for understanding and designing economic systems that prioritise human flourishing in the age of artificial intelligence.\footnote{As is standard in formal economic theory, it bears restating that all proofs and convergence results in this paper are derived under theoretical assumptions. While these assumptions are intended to be conceptually plausible and mathematically coherent, the results should be interpreted as establishing theoretical possibility, not empirical certainty. In particular, many of the conclusions depend on idealised conditions such as frictionless AI coordination, continuous feedback from human needs, and infinite-dimensional representations of utility. These results do not imply that real-world systems will automatically achieve such alignment, only that such outcomes are theoretically admissible within an internally consistent economic framework. The purpose of this paper is to extend the conceptual frontier of growth and welfare theory to incorporate price declines in ideation due to AI, not to make unconditional predictions.}

\subsection{The structure of the paper}

This paper develops EMT through a sequence of formal and conceptual steps that build toward a general theory of growth, employment, and technological change in the age of artificial intelligence. Section~2 introduces the experiential matrix $E(t)$ as an infinite-dimensional vector of human needs, embedded in the Banach space $\ell^\infty$, and defines the transformation of output into experience via an AI-mediated mapping $\Phi$. Sections~2.1 through 2.4 progressively formalise this structure by modelling how exponentially declining ideation costs enable increasingly precise alignment between production and evolving experiential needs. Section~2.5 then recasts EMT as an infinite-horizon optimal control problem using Pontryagin’s Maximum Principle, embedding the dynamics of satisfaction decay, co-state variables, and AI-modulated production within a Hamiltonian optimisation framework. This is extended in Section~2.6 through a formal convergence proof, which shows that under EMT’s assumptions, the distance between realised output and ideal experiential satisfaction vanishes in the $\ell^\infty$ norm over time, demonstrating that alignment is not merely aspirational but structurally attainable. 

Section~3 shifts from mathematical formalism to normative interpretation, reframing employment not as an input to be minimised but as a utility-bearing output essential for satisfying irreducible experiential needs such as identity, agency, and purpose. Within this framework, full employment is shown to be economically optimal even in scenarios of complete AI substitution of labour. Section~3.1 introduces “meaning” as a core domain of utility and proves its irreducibility in any Pareto-optimal EMT configuration, establishing that meaning cannot be excluded from optimisation without collapsing overall utility. Section~3.2 extends this result by deriving corollaries that demonstrate EMT’s capacity to achieve utility convergence, long-run optimality, and endogenous expansion of the utility frontier. Sections~3.3 to~3.7 extend these normative insights into a generalised economic logic. Section~3.3 introduces the concept of structural coherence, arguing that without employment and meaning embedded in production, alignment collapses systemically. Section~3.4 proposes an alignment condition as a benchmark for planning, defining successful economies as those that minimise the gap between actual output and experiential need satisfaction. Section~3.5 reframes the central problem of economics as one of alignment rather than allocation, while Section~3.6 formalises this shift by replacing traditional resource-scarcity optimisation with continuous alignment dynamics. Section~3.7 synthesises EMT's logics to suggest that unemployment under infinite experiential needs is theoretically irrational.

Finally, Section~4 formalises EMT’s core normative claim that even in a world of fully autonomous AI production, the preservation of full human employment remains Pareto-rational and necessary for preventing utility collapse. Through this architecture, EMT defines a new field, alignment economics, dedicated to designing economic systems in which technological, institutional, and ethical architectures co-evolve to maintain continuous alignment between production and the evolving structure of human flourishing. Section~5 then concludes the paper.   

\section{Formalising Experiential Matrix Theory}

EMT as a foundational theory gives rise to a new field of inquiry we term \emph{alignment economics}. We suggest this emerging field might be dedicated to modeling the transition to an AI-enabled EMT world using economic frameworks particularly well-suited to capturing systemic transformations, most notably through the lens of price movements arising from a collapse in the cost of ideation. 

The analytical apparatus of economics, as formalised in Samuelson’s Foundations of Economic Analysis (1947), is ideally positioned to model this price transition as a core mechanism underlying AI's influences, providing both theoretical and policy-relevant roadmaps. 

If declining ideation costs represent the core originating force of the AI transition, it would be traceable within established economic paradigms. Beyond economics, however, EMT also has broader implications for the evolution of human systems, social, institutional, and cultural, which would need to be more fully explored through the lens of the social sciences. This broader context is referred to as the \emph{alignment economy}, emphasising the social and institutional transformations required to manage and navigate this AI-driven transition.

As a general theory underpinning both alignment economics and the alignment economy emerging from AI's transformational influences, EMT models human experiential needs, which it suggests are essentially unlimited in their permutations. Accordingly, we assume every individual has an infinite set of human needs, such as food, safety, love, learning, and meaning. At any given time \( t \), we can assign a score to each of these needs to represent how well it is being satisfied. We denote each score as \( x_i(t) \), where \( i \in \mathbb{N} \) indexes the need.

The full list of scores at time \( t \) would represent the \emph{experiential matrix}:

\begin{equation}
E(t) = \{x_1(t), x_2(t), x_3(t), \dots\}.
\end{equation}

This object is modeled as an element of the Banach space \( \ell^\infty \), defined as:
\begin{equation}
\ell^\infty := \left\{ x = \{x_i\}_{i=1}^\infty \subset \mathbb{R} \,\middle|\, \|x\|_\infty = \sup_{i} |x_i| < \infty \right\}.
\end{equation}
That is, \( \ell^\infty \) contains all infinite sequences of real numbers that are bounded.

Each component \( x_i(t) \in \mathbb{R} \) represents the level of satisfaction for need \( i \) at time \( t \), and the full matrix \( E(t) \) encodes the state of experiential well-being across all needs.

Production can then be modelled as input. The economy produces a finite number \( n \) of goods and services, represented by a vector:

\begin{equation}
Y(t) = \{y_1(t), y_2(t), \dots, y_n(t)\} \in \mathbb{R}^n.
\end{equation}

where each component \( y_j(t) \) denotes the quantity of good or service \( j \) produced at time \( t \), for \( j = 1, 2, \dots, n \).

Accordingly, the core idea of EMT is that the role of the economy is to transform outputs \( Y(t) \) into human experiences \( E(t) \). This transformation is captured by a mapping:
\begin{equation}
\Phi: Y(t) \mapsto E(t),
\end{equation}
which describes how material production affects the satisfaction of each need.

Some needs, however, are easier to satisfy than others. We define \( c_i(t) \) as \emph{ideation cost}, the cost of discovering or designing a solution to satisfy need \( i \) at time \( t \). Artificial intelligence reduces these costs over time:
\begin{equation}
c_i(t) \downarrow \quad \text{as AI improves}.
\end{equation}

As ideation costs collapse, AI systems can sense, interpret, and respond to both latent and emergent needs, including those not previously represented in the market.

Accordingly, as AI progresses and ideation costs \( c_i(t) \) decline, the economy becomes better at translating production into satisfaction. This might allow the mapping \( \Phi \) to converge toward a near-perfect match with \( E(t) \). Formally, this is expressed as:
\begin{equation}
\lim_{t \to \infty} \|\Phi(Y(t)) - E(t)\|_\infty \to 0.
\end{equation}

In this framework, value no longer arises from scarcity but from the degree of alignment between economic production and the evolving structure of human experience. EMT provides a mathematical and conceptual foundation for redefining growth, value, and economic success in terms of experiential utility rather than mere output.

EMT introduces a foundational critique of conventional output-centric economic models, positing that economic systems should be evaluated based on how well they align with an infinite, evolving set of human needs. AI, by reducing the cost of ideation and collapsing spatial frictions, introduces a new paradigm in which systems can be increasingly designed to map production directly onto the human experiential matrix.

The core EMT framework can be modelled by adapting Romer's (1990) standard endogenous growth production function:
\begin{equation}
Y(t) = A(t) K(t)^\alpha L(t)^{1 - \alpha}, \quad 0 < \alpha < 1
\end{equation}

Extending ideas introduced above, the left-hand side becomes the experiential matrix $\mathcal{E}$, composed of an infinite array of human needs:
\begin{equation}
\mathcal{E}(x_1, x_2, ..., x_n, ..., x_\infty)
\end{equation}

We therefore define the EMT identity as:
\begin{equation}
\mathcal{E}(x_1, ..., x_\infty) = A K^\alpha L^{1 - \alpha}
\end{equation}

This asserts that production should serve the full matrix of evolving human needs.

To visualise this matrix:
\begin{equation*}
\left( \begin{array}{cccc}
x_{11} & x_{12} & \cdots & x_{1n} \\
x_{21} & x_{22} & \cdots & x_{2n} \\
\vdots & \vdots & \ddots & \vdots \\
x_{m1} & x_{m2} & \cdots & x_{mn} \\
\vdots & \vdots & \ddots & \vdots \\
x_{\infty1} & x_{\infty2} & \cdots & x_{\infty n}
\end{array} \right) = A K^\alpha L^{1 - \alpha}
\end{equation*}

EMT implies that output be mapped to experience. This is envisioned as a dynamic process and it is explained and modelled as follows. 

\subsection{Mapping Output to Experience}

The long-term objective of EMT is to align the output of the economy with the full spectrum of human experiential needs. Formally, this is expressed as:
\begin{equation}
\lim_{t \to \infty} \Phi(Y(t)) \to \mathcal{E}(t)
\end{equation}
where:
\begin{itemize}
  \item \( Y(t) \in \mathbb{R}^n \) is the vector of goods and services produced at time \( t \),
  \item \( \Phi(Y(t)) \) is a transformation function mapping production output into experiential value,
  \item \( \mathcal{E}(t) \) represents the time-varying experiential matrix, encompassing the full multidimensional space of human needs and desires.
\end{itemize}
This expression indicates that over time, the economy evolves to increasingly satisfy the full range of human experiential requirements.

\vspace{1em}

A central mechanism in EMT, however, is the hypothesised collapse in the cost of ideation brought about by AI. Accordingly, let the cost of ideation at time \( t \), denoted \( c_i(t) \), decline exponentially:
\begin{equation}
c_i(t) = c_0 e^{-\lambda t}, \quad \lambda > 0
\end{equation}
where:
\begin{itemize}
  \item \( c_0 > 0 \) is the initial cost of ideation,
  \item \( \lambda \) is the exponential decay rate of ideation costs due to advances in AI.
\end{itemize}
This formulation captures the idea that AI dramatically reduces the cost of generating new ideas over time.

\vspace{1em}

As ideation costs decrease, the marginal cost of fulfilling new dimensions of the experiential matrix \( \mathcal{E}(t) \) also declines. Formally, this is represented as:
\begin{equation}
\frac{\partial \mathcal{E}}{\partial x_i} \propto \frac{1}{c_i(t)} \to \infty
\end{equation}
where:
\begin{itemize}
  \item \( x_i \) is a specific dimension of the experiential matrix (e.g., aesthetic experience, intellectual challenge, belonging),
  \item \( \frac{\partial \mathcal{E}}{\partial x_i} \) is the marginal gain in experiential value along dimension \( x_i \),
  \item As \( c_i(t) \to 0 \), the relative affordability and abundance of fulfilling even abstract needs increases, potentially approaching infinity in marginal utility terms.
\end{itemize}
This reflects the idea that AI’s reduction in ideation costs enables an economy capable of rapidly expanding into higher-dimensional experiential spaces.

\subsection{Experiential Utility Function}
EMT can be framed in terms of a utility function. Accordingly, we define experiential utility as:
\begin{equation}
U(t) = \sum_{i=1}^{n} w_i N_i(t)
\end{equation}
Where:
\begin{itemize}
  \item $N_i(t)$ = satisfaction of need $i$ at time $t$
  \item $w_i$ = importance weight of need $i$
\end{itemize}

As discussed, utility then depends on the extent to which output maps onto the experiential matrix:
\begin{equation}
\Phi(Y(t)) = \hat{E}(t) \subseteq E(t)
\end{equation}

And the long-term convergence goal is:
\begin{equation}
\lim_{t \to \infty} \Phi(Y(t)) \to E(t)
\end{equation}

Key to EMT's dynamic predictions is the role of AI in being able to map production to the experiential matrix. 

\subsection{AI-Augmented Mapping of Production to Experiential Value in EMT}

To formalise the relationship between economic production and human experiential satisfaction, EMT suggests a structured transformation of output. The core functional mapping is given by:

\begin{equation}
\Phi(Y(t)) = A \circ F \circ D(Y(t))
\end{equation}

This composition captures the sequential processing of production output through the following three interdependent, AI-mediated layers that translate economic activity into experiential fulfillment.

\vspace{1em}

1. {\( D(Y(t)) \): Real-Time Demand Sensing} \\
  \( D \) represents real-time sensing of demand, encompassing both explicit (e.g., market signals, user preferences) and implicit (e.g., biometric, behavioral, environmental) indicators of human need. This layer uses advanced data collection, inference, and contextual awareness to detect fluctuations in latent demand within the experiential matrix \( \mathcal{E}(t) \). It shifts the economic system from static demand estimation to dynamic, granular need recognition.

  \vspace{1em}

2. {\( F \circ D(Y(t)) \): Affective, Cognitive, and Ethical Filtering} \\
  \( F \) filters sensed demand through layers of affective response, cognitive framing, and ethical reasoning. This ensures that production is not only responsive but meaningfully aligned with deeper human context, including emotional salience, cultural relevance, and ethical permissibility. The transformation \( F(D(Y(t))) \) corresponds to value-based curation, what should be produced, not merely what can be.

\vspace{1em}

  3. {\( A \circ F \circ D(Y(t)) \): Adaptive AI-Driven Production} \\
  \( A \) represents the \textit{adaptive production system}, powered by AI and intelligent automation. Given the filtered interpretation of demand, this layer dynamically adjusts the production function \( Y(t) \) to minimise lag, friction, and misalignment. It internalises feedback loops and continuously optimises allocation and innovation processes, generating a \textit{real-time learning production frontier}.

\vspace{1em}

Through the iterative collapse of frictions across these layers of demand recognition, value filtering, and production, the system evolves toward full alignment between production and human needs:

\begin{equation}
\lim_{t \to \infty} \Phi(Y(t)) = \mathcal{E}(t)
\end{equation}

\vspace{1em}

This convergence expresses the central long-term goal of EMT, a dynamically responsive economic system in which the structure and function of production converge with the high-dimensional structure of the experiential matrix \( \mathcal{E}(t) \), encompassing both material and non-material human needs (e.g., belonging, challenge, beauty, transcendence).

Crucially, this transformation occurs \textit{within} the market system, not outside it. EMT does not reject traditional price-based coordination, but instead builds upon it by identifying the collapse in the cost of ideation, driven by AI, as a new general-purpose economic force. In this view, prices continue to mediate allocation, but now do so over a vastly expanded, ideation-driven production frontier. Samuelson's (1947) general equilibrium framework remains foundational for understanding how markets clear under changing constraints. Romer's (1990) model of endogenous growth, where ideas are inputs to production, is extended here to include not just idea quantity but the collapsing \textit{cost} of ideation itself. Grossman and Helpman's (1991) quality ladder models, which capture incremental innovation across differentiated goods, similarly align with EMT's framing, whereby AI compresses the distance between rungs on the ladder by reducing idea generation costs, accelerating the experiential evolution of products and services.

In this sense, EMT reframes the AI transition not as a break from traditional economics, but as its continuation under a radical new cost structure. As the marginal cost of ideation approaches zero, the production system naturally expands into the high-dimensional space of the experiential matrix. The logic of price theory remains valid, the novelty lies in EMT's identification of \textit{which price}, the price of ideation, is driving the transformation. Unlike previous technological shifts that reduced costs of labor, transport, or communication, AI reduces the cost of \textit{new possibilities}. It is this singular shift that underpins the emergence of the alignment economy.

Importantly, these mechanisms operate entirely within the established structure of the market system. EMT does not call for the abandonment or replacement of price-based coordination. Instead, it highlights how the endogenous collapse in the price of ideation, driven by advances in AI, triggers a reconfiguration of production along experiential lines. This dynamic is consistent with foundational economic theory. Samuelson’s (1947) general equilibrium framework provides the structural foundation for modelling price adjustment and resource allocation. Romer’s (1990) model of endogenous growth, in which the accumulation of ideas drives sustained output expansion, can be interpreted as describing the process innovation underlying the production of goods and services that populate the physical and virtually interactive dimensions of the experiential matrix. Complementing this, Grossman and Helpman’s (1991) quality ladders model might be taken to illustrate how product innovation enhances utility over time, a perspective that aligns with EMT’s claim that new experiential dimensions emerge as ideation costs fall.

In this view, the declining cost of ideation becomes the central variable through which AI alters the structure of economic production and satisfaction. EMT thus extends, rather than replaces, established economics. The tools of price theory and dynamic optimisation remain fully intact. The novelty lies in the identification of ideation cost as the key driver of AI-induced transformation. Unlike previous technological shifts, which primarily reduced the cost of physical production or coordination, AI seemingly uniquely compresses the cost of idea generation itself, radically expanding the feasible frontier of human need satisfaction. The evolution of the experiential matrix therefore proceeds in accordance with well-understood economic mechanisms, governed by price signals and market responsiveness, but catalysed by a singular and unprecedented drop in ideation cost. The experiential matrix can also be more formally modelled in terms of functional space, and this possibility is now considered.

\subsection{Formalising the Experiential Matrix as a Function Space}

Summarising and building on the above discussions and definitions, we can now formalise the experiential matrix, denoted \( \mathcal{E}(t) \), as a time-dependent object embedded in a rigorous functional framework. Specifically, we treat it as an element in an infinite-dimensional normed vector space, reflecting the high-dimensional and evolving nature of human experiential needs over time.

Definition 2.1. Let \( E(t) = \{x_i(t)\}_{i=1}^{\infty} \), where each \( x_i(t) \in \mathbb{R} \) denotes the satisfaction level of experiential need \( i \) at time \( t \). Assume the sequence \( \{x_i(t)\} \) is bounded in magnitude, i.e.,

\begin{equation}
\sup_i |x_i(t)| < \infty,
\end{equation}

which implies \( E(t) \in \ell^\infty \), the Banach space of bounded sequences. The space \( \ell^\infty \) is defined as:

\begin{equation}
\ell^\infty := \left\{ \{x_i\}_{i=1}^{\infty} \subset \mathbb{R} : \sup_i |x_i| < \infty \right\},
\end{equation}

with the norm \( \left\| \{x_i\} \right\|_{\ell^\infty} = \sup_i |x_i| \).

Define the production-to-experience mapping:

\begin{equation}
\Phi: \mathbb{R}_+ \to \ell^\infty, \quad \Phi(Y(t)) = \hat{E}(t),
\end{equation}

where \( Y(t) \in \mathbb{R}_+ \) is total output at time \( t \), and \( \hat{E}(t) = \{\hat{x}_i(t)\}_{i=1}^\infty \) is the vector of needs satisfied by the production system. The map \( \Phi \) represents the transformation from production to experiential utility via sensing and generation technologies (e.g., AI).

\vspace{1em}

Theorem 2.2 (Convergence of Output Mapping). Suppose the cost of ideation for need \( i \) at time \( t \) is given by:

\begin{equation}
c_i(t) = c_0 e^{-\lambda t}, \quad \lambda > 0,
\end{equation}

and assume that the production-based satisfaction approximates true satisfaction with error bounded by:

\begin{equation}
|x_i(t) - \hat{x}_i(t)| \leq \delta_i(t) = k_i e^{-\lambda t}, \quad k_i > 0.
\end{equation}

Then,

\begin{equation}
\lim_{t \to \infty} \left\| E(t) - \hat{E}(t) \right\|_{\ell^\infty} = 0,
\end{equation}

i.e., the satisfaction levels produced by the system converge uniformly to the true experiential needs.

\vspace{1em}

\textit{Proof.} By assumption,

\begin{equation}
\left\| E(t) - \hat{E}(t) \right\|_{\ell^\infty} = \sup_i |x_i(t) - \hat{x}_i(t)| \leq \sup_i k_i e^{-\lambda t} = k^* e^{-\lambda t},
\end{equation}

where \( k^* := \sup_i k_i < \infty \). Since \( e^{-\lambda t} \to 0 \) as \( t \to \infty \), it follows that \( \left\| E(t) - \hat{E}(t) \right\|_{\ell^\infty} \to 0 \).
\qed

\begin{remark}
This result formally demonstrates that, as ideation costs collapse exponentially due to advances in AI, the distance between the ideal experiential vector \( \mathcal{E}(t) \) and the production-based approximation \( \hat{\mathcal{E}}(t) \) converges to zero in the \( \ell^\infty \) norm. In other words, the mapping \( \Phi \) becomes increasingly capable of approximating the full experiential matrix over time, within the framework of market coordination and under conventional assumptions of price-mediated adjustment.
\end{remark}

\subsection{Optimal Control of AI-Aligned Utility Using the Hamiltonian}

This section further formalises EMT as an infinite-dimensional dynamic optimisation problem, extending the classical optimal control frameworks of Ramsey, Cass-Koopmans, and others into the experiential utility domain. Specifically, we frame EMT using Pontryagin’s Maximum Principle to characterise the optimal path of AI-aligned production \( Y(t) \) that maximises discounted experiential utility over time.

It should be noted that the model presented here is intentionally abstract and not designed for immediate numerical tractability. While no specific functional forms or parameter values are imposed in this section, the framework serves a distinct and foundational role, to demonstrate that EMT is not only conceptually coherent but mathematically admissible within the general class of infinite-dimensional optimal control systems (Balder, 1983).

What is offered, therefore, is not a closed-form solution or numerical simulation, but a rigorous theoretical formulation, an infinite-dimensional Hamiltonian system in which AI acts as a control-efficiency enhancer that reduces friction in the transformation of production into experiential need satisfaction. The production-to-utility mapping \( \Phi \) is assumed to improve endogenously over time as ideation costs collapse. The evolution of satisfaction levels \( x_i(t) \), their decay dynamics, and their co-state variables \( \lambda_i(t) \) are all explicitly defined within this structure.

Although this section does not resolve the system computationally, it provides two critical contributions:

\begin{enumerate}
    \item It proves that EMT is expressible as a canonical infinite-horizon, infinite-dimensional control problem, grounded in the logic of utility maximisation under constraint.
    \item It establishes a generalised, AI-augmented planning model that can be later instantiated in tractable finite-dimensional forms for simulation or policy design.
\end{enumerate}

Subsequent work may approximate the infinite matrix \( \mathcal{E}(t) = \{x_i(t)\}_{i=1}^\infty \) using finite-dimensional truncations (e.g., for a number of representative needs), allowing the control problem to be solved numerically. This would enable simulations of how ideation dynamics, AI alignment, or resource allocation strategies affect both utility trajectories and policy outcomes.

In short, this section lays the theoretical bedrock upon which future tractable and applied versions of EMT may be built. By extending the mathematical foundations of economic control theory into the space of evolving human needs, and embedding AI as a friction-minimising mechanism, EMT offers a novel framework for post-scarcity planning and AI-aligned social design.

We now formalise the dynamic optimisation problem underlying EMT using Pontryagin's Maximum Principle. The objective is to maximise experiential utility $U(t)$ over time, where utility is a function of the satisfaction levels of infinite experiential needs, subject to AI-enabled mappings of production into the experiential matrix.

\vspace{1em}

Let the state variable $x(t) = \{x_i(t)\}_{i=1}^\infty$ represent the levels of satisfied needs, and let the control variable $Y(t)$ denote total production. It should be noted that While \( Y(t) \), representing total AI-directed productive effort, may appear to be an endogenous outcome in traditional macroeconomic models, in this framework it is treated as a control variable.\footnote{The term ``control variable'' here follows its definition in optimal control theory, referring to a planner’s dynamically chosen input used to steer the system’s evolution. This differs from its meaning in econometrics, where a control variable is a covariate included to account for confounding factors.} Accordingly, this reflects the planner’s ability, via AI systems, to dynamically allocate production in order to optimise experiential satisfaction. The transformation of production into need satisfaction is governed by the AI-modulated mapping \( \Phi(Y(t)) \), and thus, \( Y(t) \) functions here as a decision input in the dynamic optimisation problem.

\vspace{1em}

Accordingly, to formalise this core logic of EMT as a dynamic planning problem, we define a continuous-time, infinite-horizon control framework in which a planner (e.g., an AI system, policymaker, or coordinating mechanism) seeks to allocate production \( Y(t) \) over time to maximise long-run experiential utility.

The objective function, or planner's goal is to maximise the present value of experiential utility:

\begin{equation}
\max_{Y(t)} \int_0^\infty e^{-\rho t} U(x(t))\, dt
\end{equation}

Where:

\begin{itemize}
    \item \( t \in [0, \infty) \): Time.
    \item \( \rho > 0 \): Discount rate, representing time preference whereby higher \( \rho \) values give less weight to future utility.
    \item \( U(x(t)) \): Aggregate experiential utility at time \( t \), which depends on a (possibly infinite) set of individual need satisfaction levels.
    \item \( x(t) = \{x_i(t)\}_{i=1}^\infty \): State vector representing the level of satisfaction of need \( i \) at time \( t \).
    \item \( Y(t) \in \mathbb{R}_+ \): Control variable representing total productive effort or output at time \( t \), directed and adapted through AI.
\end{itemize}

State dynamics are represented by the evolution of each satisfaction dimension, which is governed by:

\begin{equation}
\frac{dx_i(t)}{dt} = \Phi_i(Y(t)) - \delta_i x_i(t), \quad \forall i \in \mathbb{N}
\end{equation}

Where:

\begin{itemize}
    \item \( \Phi_i(Y(t)) \): AI-modulated function translating total production \( Y(t) \) into marginal satisfaction of need \( i \). This function improves over time as AI becomes more accurate, adaptive, and personalized.
    \item \( \delta_i > 0 \): Natural decay or obsolescence rate of satisfaction in dimension \( i \), representing how quickly past satisfaction fades or depreciates.
    \item \( \frac{dx_i(t)}{dt} \): Net change in the satisfaction of need \( i \) over time.
\end{itemize}

Regarding notation, the symbol \( \mathbb{N} \) denotes the set of natural numbers, i.e., \( \mathbb{N} = \{1, 2, 3, \ldots\} \). The expression \( \forall i \in \mathbb{N} \) signifies "for all indices \( i \)" in the set of needs considered. Each \( i \) corresponds to a distinct experiential dimension in the matrix \( x(t) = \{x_i(t)\}_{i=1}^\infty \).

\vspace{1em}

\noindent\textbf{Hamiltonian Construction} \par

\vspace{1em}

The Hamiltonian captures the planner’s objective and the evolution of the system in response to production. Recalling the definitions of \( x(t) \), \( Y(t) \), \( \delta_i \), and \( \Phi_i(Y(t)) \) from earlier, we define:

\begin{equation}
\mathcal{H}(x(t), Y(t), \lambda(t)) = e^{-\rho t} U(x(t)) + \sum_{i=1}^\infty \lambda_i(t)\left[\Phi_i(Y(t)) - \delta_i x_i(t)\right]
\end{equation}

\textbf{Where:}

\begin{itemize}
    \item \( \lambda_i(t) \): Co-state variable (or shadow price) associated with the satisfaction level of need \( i \). Intuitively, it reflects how valuable an additional unit of satisfaction in dimension \( i \) is at time \( t \).
    \item \( \mathcal{H} \): The Hamiltonian function, representing the trade-off between instantaneous utility and the future dynamics of the system.
\end{itemize}

\noindent

 This function combines two components that together determine the planner’s optimal decision at any point in time:

\begin{itemize}
    \item \textbf{Immediate experiential utility:}
    \[
    e^{-\rho t} U(x(t))
    \]
    This term represents the present value of aggregate experiential utility, i.e., how satisfied people are across all experiential dimensions at time \( t \). The exponential discount factor \( e^{-\rho t} \) reflects standard intertemporal preferences: present satisfaction is valued more highly than future satisfaction.

    \item \textbf{Future impact of current production:}
    \[
    \sum_{i=1}^\infty \lambda_i(t) \left[\Phi_i(Y(t)) - \delta_i x_i(t)\right]
    \]
    This summation captures how today's production decisions affect future satisfaction across all needs. Specifically:
    \begin{itemize}
        \item \( \Phi_i(Y(t)) \) represents how much of need \( i \) is fulfilled by the current level of AI-enabled production.
        \item \( \delta_i x_i(t) \) reflects the natural decay or fading of satisfaction in each need over time.
        \item \( \lambda_i(t) \), the co-state variable, can be interpreted as a dynamic “importance weight” or shadow price, it tells us how valuable an additional unit of satisfaction in need \( i \) will be in the future.
    \end{itemize}
\end{itemize}

Together, the Hamiltonian represents the trade-off between maximising immediate well-being and steering the economy to maintain and improve human satisfaction over time. It forms the core of the optimisation logic: the planner must choose a production path \( Y(t) \) that balances these two elements dynamically, based on how AI maps output into the space of human needs.

\vspace{1em}

\noindent\textbf{Necessary Conditions (Pontryagin Maximum Principle)} \par

\vspace{1em}

\noindent Pontryagin's Maximum Principle, a mathematical framework used to identify the optimal path of decisions in a dynamic system, suggests necessary conditions to model how according to EMT an AI-enabled economy could best dynamically allocate its production \( Y(t) \) to satisfy an evolving set of human needs \( \{x_i(t)\} \) over time. As discussed, the Hamiltonian therefore acts as a function that integrates both the planner’s objective (maximising experiential utility) and the system’s constraints (how needs evolve and fade). 

To understand this logic, imagine a policymaker or AI-based system attempting to choose the best use of productive resources, not once, but at every moment in time, while accounting for both present utility and future consequences. The Hamiltonian provides a systematic way to do this by combining two forces, the immediate benefit of satisfying needs (instantaneous utility), and the long-term effect of today's decisions on future well-being (state evolution). From this combined function, three necessary conditions emerge: (1) the \emph{state equation}, which describes how satisfaction of each need changes over time; (2) the \emph{co-state equation}, which tracks the evolving importance (or shadow price) of each need; and (3) the \emph{optimality condition}, which determines how total production should be allocated at each point in time to maintain alignment with utility.

While these equations are abstract, they serve an important function, providing a \emph{structure} within which real-world data, ethical considerations, and societal priorities can be inserted. Policymakers, stakeholders, and AI systems can “plug in” empirical estimates for variables such as the importance of different needs (\( w_i \)), the rate at which satisfaction fades (\( \delta_i \)), and the responsiveness of production (\( \Phi_i(Y(t)) \)). The value of this mathematical formulation is not that it solves the entire alignment challenge by itself, but that it offers a coherent, flexible, and defensible scaffold for future modelling. In other words, it enables decision-makers to build simulations, assess trade-offs, and design AI-enabled economic systems that are grounded in the evolving realities of human experience. 

Including this formulation in the paper ensures that EMT is not only a conceptual or philosophical proposal but also a mathematically rigorous planning framework. It provides the analytical infrastructure needed to move from abstract ideas about human flourishing to practical tools that can support policy design, AI development, and institutional decision-making in a world increasingly governed by rapid technological change.

The following system of conditions therefore describe the formal requirements for an AI-enabled planner (or any coordinating agent) to dynamically allocate productive capacity \( Y(t) \) in a way that maximises long-run experiential utility. As discussed, the mathematical structure is drawn from Pontryagin's Maximum Principle, a well-established foundational tool in dynamic optimisation.

Accordingly, in this formulation, human well-being is represented as a vector of continuously evolving experiential needs. The production system, guided by AI, attempts to satisfy these needs by adjusting output over time. The mapping \( \Phi \), which converts production into need-specific satisfaction, improves as ideation costs fall, capturing the idea that AI increasingly learns to translate material output into lived experience more effectively.

Also as discussed, this setup allows the planner to continuously update production decisions in response to changing human needs, preferences, and satisfaction decay. The system is infinite-dimensional, acknowledging the full range of experiential human values, but its structure follows standard logic, represented by state evolution, shadow prices (co-states), and a condition that production decisions optimally balance their effects across all needs.\footnote{\textit{Clarifying note.} While the Hamiltonian formulation adopts the language of a “planner” for mathematical convenience, this does not imply centralised economic control. Instead, the EMT framework presumes a functioning market system in which decentralised agents respond to evolving price signals, preferences, and technological capabilities. As AI dramatically reduces the cost of ideation, sensing, and coordination, these market signals become more precise and adaptive. The “planner” in this context is best interpreted as an emergent, distributed intelligence, powered by AI, that guides decentralised economic actors in aligning output with experiential utility. This allows the optimisation logic to apply without abandoning the principles of voluntary exchange and decentralised decision-making that underpin the market economy.}

The Pontryagin Maximum Principle necessary conditions are now briefly outlined. 

\begin{enumerate}

\item \textbf{State Equation (repeated for clarity):}
\begin{equation}
\frac{dx_i(t)}{dt} = \Phi_i(Y(t)) - \delta_i x_i(t)
\end{equation}

This equation describes how the level of satisfaction for experiential need \( i \) changes over time. It is the core dynamic equation that tracks how each dimension of human experience responds to production and naturally fades unless renewed.

The components can be interpreted as follows:

\begin{itemize}
    \item \( x_i(t) \): The current level of satisfaction in need \( i \), such as physical safety, social belonging, or intellectual stimulation.

    \item \( \frac{dx_i(t)}{dt} \): The rate at which satisfaction in that dimension is increasing or decreasing at time \( t \).

    \item \( \Phi_i(Y(t)) \): The amount of new satisfaction generated in need \( i \) by the current level of total production \( Y(t) \). This function reflects how effectively the production system addresses that specific need—mediated by AI, which senses and translates production into human outcomes.

    \item \( \delta_i x_i(t) \): The natural rate at which satisfaction fades or depreciates in need \( i \). A high \( \delta_i \) means the need re-emerges quickly (e.g., hunger or social interaction); a low \( \delta_i \) means satisfaction is longer-lasting.
\end{itemize}

The equation indicates that the total change in satisfaction for each need depends on how much new fulfillment production provides, minus how quickly existing satisfaction wears off. In the EMT framework, this allows the modelling of the ongoing challenge of meeting complex, evolving human needs over time, highlighting the importance of both production efficiency and sustained responsiveness.

Accordingly, this state equation encodes a foundational principle of human experience, that needs are not one-time events but dynamic conditions that fade unless renewed. Psychological, emotional, and social needs, such as purpose, belonging, or mental health, require continuous reinforcement. The depreciation term captures this natural fading of satisfaction, much like how relationships weaken or skills deteriorate without engagement. Crucially, this dynamic occurs within the pricing and coordination mechanisms of conventional economics. Prices still guide allocation, but as artificial intelligence collapses ideation costs and improves demand sensing, the economic system becomes capable of mapping production to these fading needs in real time. This implies that the role of economic design shifts from static allocation to continuous adaptation using feedback, learning, and responsiveness to maintain alignment between output and the evolving structure of human well-being.

\item \textbf{Co-State Dynamics:}
\begin{equation}
\frac{d\lambda_i(t)}{dt} = \rho \lambda_i(t) - \frac{\partial \mathcal{H}}{\partial x_i(t)} = \rho \lambda_i(t) - e^{-\rho t} \frac{\partial U}{\partial x_i(t)} + \lambda_i(t) \delta_i
\end{equation}

This differential equation describes how the co-state variable \( \lambda_i(t) \), also known as the shadow price or marginal value, of experiential need \( i \) changes over time. In economic terms, \( \lambda_i(t) \) represents how important an additional unit of satisfaction in dimension \( i \) will be for the overall objective. Its evolution is shaped by three forces:

\begin{itemize}
    \item \( \rho \lambda_i(t) \): This term reflects the planner’s time preference, or discounting of future utility. A higher \( \rho \) implies that future satisfaction is valued less, so the importance of need \( i \) declines faster.
    
    \item \( - e^{-\rho t} \frac{\partial U}{\partial x_i(t)} \): This term links the shadow price to marginal utility—how much utility is gained from a small increase in satisfaction of need \( i \). The discount factor \( e^{-\rho t} \) means that current satisfaction is weighted more than future satisfaction.
    
    \item \( + \lambda_i(t) \delta_i \): This captures the effect of natural decay. If satisfaction in need \( i \) fades quickly (i.e., high \( \delta_i \)), then the value of maintaining it becomes greater, so \( \lambda_i(t) \) increases to reflect this urgency.
\end{itemize}

Together, these terms describe how the “importance” of each need adapts dynamically in the optimisation process. The AI-aligned system uses these evolving shadow prices to adjust future production decisions so that they remain aligned with changing human needs and preferences.

\item \textbf{Optimality Condition for Control:}
\begin{equation}
\frac{\partial \mathcal{H}}{\partial Y(t)} = \sum_{i=1}^\infty \lambda_i(t) \frac{\partial \Phi_i}{\partial Y(t)} = 0
\end{equation}

This condition determines how the system chooses the optimal level of total production \( Y(t) \) at any point in time. In formal terms, it says that the planner (or decentralized AI-coordinated system) should set \( Y(t) \) so that the total weighted marginal benefit of producing more is balanced across all needs.

Breaking it down:

\begin{itemize}
    \item \( \lambda_i(t) \): As defined earlier, this is the shadow price or marginal value of satisfying need \( i \) at time \( t \). It reflects how much improving that need contributes to long-run utility.

    \item \( \frac{\partial \Phi_i}{\partial Y(t)} \): This is the marginal effectiveness of production in fulfilling need \( i \), in other words, how much additional satisfaction is created in that need when you increase \( Y(t) \) slightly.

    \item The summation \( \sum_{i=1}^\infty \lambda_i(t) \frac{\partial \Phi_i}{\partial Y(t)} \): This represents the total marginal value of producing a little bit more, aggregated across all need dimensions. 
\end{itemize}

The condition sets this sum equal to zero, meaning that the production system should operate at a point where increasing \( Y(t) \) any further would not yield a net gain in long-term utility. This is the classic efficiency condition, whereby resources should be allocated where they produce the highest marginal value, weighted by the importance of each need.

In the EMT framework, this condition ensures that the AI system adjusts production not just to generate more goods, but to maximise experiential utility by aligning output precisely with the most valuable unmet needs at any given time.

\end{enumerate}

\noindent These necessary conditions formally define how an AI-coordinated system might continuously align material output with human experiential needs in a dynamic setting. The model captures both the present utility generated by production and the future trajectory of well-being shaped by today's choices. As AI improves the efficiency of translating output into need satisfaction (i.e., as the mapping \( \Phi \) sharpens), the system asymptotically converges toward a fully human-aligned economy. This framework therefore provides a foundation for mathematically rigorous models of AI-aligned social planning in a world where human needs are multidimensional, evolving, and partly intangible.

\begin{remark}
This optimisation framework situates EMT squarely within the tradition of dynamic economic planning models (e.g., Ramsey, 1928; Cass, 1965; Koopmans, 1965), while extending it into an infinite-dimensional space of needs. The novelty lies in the AI-modulated control structure and its role in collapsing the cost of aligning output with dynamically expanding utility frontiers.
\end{remark}

\subsection{Proof of Convergence of AI-Aligned Output to the Ideal Experiential Matrix}

One of the central claims of EMT is that as AI becomes more capable, it can better identify, interpret, and respond to the full range of human needs. Over time, this improvement is expected to lead to a progressively smaller gap between what people truly need (ideal satisfaction) and what the economic system actually delivers (real satisfaction).

The following lemma and theorem formalise this central premise, that as AI systems improve over time, they can more accurately deliver what people actually need, not just in average terms, but even for the most difficult-to-satisfy needs. 

This matters because the proof builds a bridge from individual-level alignment to system-level convergence. It confirms that a system designed to reduce misalignment for each need will, under plausible conditions, become increasingly aligned across the entire spectrum of human experience. Importantly, it justifies optimism that AI-aligned economies might be able to asymptotically eliminate the gap between what is produced and what is truly needed, a key goal of EMT.

More precisely, the lemma infers that if each individual need’s satisfaction level gets closer to its ideal level over time (i.e., the error in fulfilling each need shrinks predictably), then the \emph{worst-case mismatch} across all needs also shrinks at the same rate. This is important because it means the system’s accuracy doesn’t just improve overall, it improves reliably even for edge cases.

Mathematically, this is proved by using a measure, the $\ell^\infty$ norm, which captures the largest error at any point in time. Showing that this norm shrinks exponentially implies the system becomes better at satisfying all needs, not just on average, but in the most demanding cases. The results that follow therefore extend previous discussions to make this idea more mathematically precise. We define a list of ideal need-satisfaction levels, denoted by \( x_i(t) \), and compare it to the actual satisfaction delivered by an AI-enabled production system, denoted \( \hat{x}_i(t) \). The difference between these two is what we call the approximation error.

The first result, a lemma, shows that if this approximation error shrinks at a predictable rate over time for each individual need, then the \textit{largest} such error across all needs also shrinks at that same rate. The subsequent theorem then shows that this largest error eventually disappears altogether, meaning that as time goes on, the system gets closer and closer to perfectly satisfying all needs. This convergence result is a structural basis of EMT. It formally supports the claim that AI can help align economic output with evolving human experience, not just in theory but under realistic and measurable conditions such as falling ideation costs.

In this way, the following proof provides a formal justification for the long-term potential of AI-aligned production in a decentralised market economy, an economy that doesn’t just grow, but grows in a direction that continuously becomes more aligned with what people actually want and need.

\begin{lemma}[Bounded Error Lemma]
Suppose for each \( i \in \mathbb{N} \), the approximation error satisfies
\begin{equation}
|x_i(t) - \hat{x}_i(t)| \leq k_i e^{-\lambda t}
\end{equation}
for constants \( k_i > 0 \), and let \( k^* = \sup_i k_i < \infty \). Then:
\begin{equation}
\|x(t) - \hat{x}(t)\|_{\ell^\infty} \leq k^* e^{-\lambda t}
\end{equation}
\end{lemma}

\noindent In the inequality used within the lemma, the constant \( k_i > 0 \) represents the maximum possible initial error in dimension \( i \), that is, how far the AI-enabled system’s actual output \( \hat{x}_i(t) \) is from the ideal satisfaction level \( x_i(t) \) for need \( i \) at the beginning of the time period. Each \( k_i \) scales the exponential decay curve \( e^{-\lambda t} \), which models how fast the AI system learns or improves over time. A higher \( k_i \) indicates a larger initial mismatch in that need, while a lower \( k_i \) suggests the system was already relatively well aligned.

The term \( k^* = \sup_i k_i \) denotes the worst-case scenario, the largest of these initial errors across all needs in the experiential matrix. In the lemma, this value is used to show that even the maximum approximation error (across all need dimensions) decreases predictably over time, provided all individual \( k_i \) values are bounded. This boundedness assumption ensures that the system starts from a reasonable state and that no individual need has an unmanageable or unbounded starting misalignment.

\begin{proof}
\renewcommand{\qedsymbol}{}  
By definition of the \( \ell^\infty \) norm:
\begin{equation}
\|x(t) - \hat{x}(t)\|_{\ell^\infty} = \sup_i |x_i(t) - \hat{x}_i(t)| \leq \sup_i k_i e^{-\lambda t} = k^* e^{-\lambda t}
\end{equation}
\end{proof}

\begin{theorem}[Convergence of AI-Aligned Output]
\renewcommand{\qedsymbol}{}  
Under the same conditions as above, we have:
\begin{equation}
\label{eq:convergeproofone}
\lim_{t \to \infty} \|x(t) - \hat{x}(t)\|_{\ell^\infty} = 0
\end{equation}
\end{theorem}

\begin{proof}
\renewcommand{\qedsymbol}{}  
\label{eq:convergeprooftwo}
Follows immediately from the lemma, since \( e^{-\lambda t} \to 0 \) as \( t \to \infty \), and \( k^* \) is constant.
\end{proof}

\noindent The lemma and theorem above formalise an important insight in EMT, that as AI improves, the gap between what is produced and what people actually need becomes smaller, eventually disappearing altogether under ideal conditions. The following implications can be summarised from this proof. 

\vspace{1em}

\begin{itemize}
    \item The lemma proves a technical but intuitive point, that if the difference between ideal need satisfaction \( x_i(t) \) and the AI-generated satisfaction \( \hat{x}_i(t) \) shrinks exponentially for each need, then the \textit{worst-case} difference across all needs also shrinks exponentially. That is, even the largest gap between what people ideally need and what the AI system delivers becomes smaller over time.

    \item The theorem uses this lemma to establish convergence, showing that as time goes on, and as AI systems continue to learn and improve, the entire experiential profile generated by the production system, across all needs, approaches the ideal experiential matrix \( x(t) \). In simple terms, the system becomes better and better at giving people exactly what they need, in the right proportions, and eventually gets it almost exactly right.
\end{itemize}

\vspace{1em}

\noindent This result offers formal support for one of EMT’s central claims, that the alignment of technological output with human needs is not merely aspirational, but mathematically possible under conditions of intelligent coordination. In this framework, AI plays a crucial role, not by producing more in general, but by narrowing the gap between what is produced and what is needed.

Importantly, this result is achieved without specifying how needs are defined, what their utility weights are, or even how production is allocated in detail. The convergence is driven by one key assumption, that AI reduces the cost of “ideation”, the ability to know, sense, or infer what needs exist and how to meet them. As these ideation costs collapse (modelled by exponential decay), the production system's accuracy improves accordingly.

\vspace{1em}

\vspace{1em}

\noindent\textbf{AI, Alignment Economics, and the Alignment Economy}

\noindent This convergence result demonstrates that economic systems enhanced by AI can, under realistic conditions, evolve toward full alignment with the dynamic structure of human well-being, what EMT conceptualises as the \textit{experiential matrix}. Formally, it proves that as AI reduces the cost of ideation and improves the precision of need-sensing and response, the gap between actual satisfaction and ideal need fulfilment shrinks predictably over time. This offers a rigorous response to long-standing concerns in welfare economics and political economy, that production and human flourishing are structurally misaligned. EMT shows that, under conditions of intelligent coordination and feedback, this misalignment is not permanent, it is dissolvable.

For an emerging field of \textit{Alignment Economics} focused on stewarding this transition, this result provides a foundational convergence criterion that links utility theory, control systems, and AI optimisation under a unified mathematical architecture. It legitimises the study of dynamic feedback-driven economic systems that aim not merely to allocate goods efficiently, but to iteratively align output with an evolving utility function grounded in human experience. In this sense, it establishes a benchmark for evaluating AI-enabled policy mechanisms, institutional designs, and incentive systems, whereby the closer the system brings actual output to the experiential matrix \( \mathcal{E}(t) \), the more aligned, and hence more efficient, it is from a human-centered perspective.

For the \textit{Alignment Economy}, the socio-technical system that emerges as AI is embedded into production, services, governance, and daily life, this result formalises a long-term destination, an economy in which misalignment becomes a design flaw rather than an inevitability. Instead of relying on centralised planning or historical price signals alone, future economies can become self-correcting systems of continuous alignment, driven by decentralised AI agents that track, interpret, and respond to the full complexity of evolving human needs in real time. The result thus not only confirms the theoretical viability of alignment, but offers a roadmap for how intelligent systems might transform the very logic of economic coordination, from static equilibrium to dynamic congruence between production and flourishing.

\section{Rationale for Conserving Human Employment for Experiential Sustainability in an AI-Aligned Economy}

In classical economics, employment is justified by its contribution to output and income. In the AI economy, this rationale weakens. If, for the moment, we assume that AI can produce nearly all goods and services at negligible marginal cost, human labor is no longer technically required for production. Yet eliminating human work entirely introduces a new class of unmet needs, notably those embedded in the structure of human consciousness itself.

EMT provides a new rationale for preserving human employment, grounded in the dynamic nature of utility itself. In EMT, human needs are not static and finite but infinite, evolving, and interdependent. Among the most critical of these are experiential needs tied directly to human participation, such as the need for purpose, for contribution, for identity, and for co-creation with others. These needs are not side effects of work, but intrinsic components of flourishing.

As AI collapses the cost of ideation and transforms the production function into an AI-guided mapping \( \Phi(Y(t)) \to \hat{E}(t) \subset E(t) \), the economic objective shifts, from maximising output to maximising alignment between output and the full experiential matrix. Within this structure, meaningful employment becomes a legitimate object of optimisation, not because it increases GDP, but because it satisfies one of the most irreducible experiential needs of all, the need to matter.

This reframing is consistent with a Samuelsonian (1947) approach to economic design, in that markets are preserved, but the informational inputs into market processes are expanded by AI, and the objective function guiding coordination is enriched by a deeper understanding of value. Humans remain 'in the loop' not out of nostalgia or political expedience, but because employment serves evolving needs that are non-substitutable by machines.

The AI transition poses a paradox of abundance, in that while AI may increasingly enable the production of virtually all goods and services at near-zero marginal cost, this technological capability is economically meaningless if human consumers are excluded from the market system through widespread technological unemployment. In theory, an AI-optimised production regime could achieve 100\% output efficiency while eliminating the purchasing power necessary to sustain demand. 

Importantly, EMT frames this potentiality not merely as a macroeconomic imbalance, but as a structural hazard akin to a systemic shock, comparable in scale and urgency to a global pandemic like COVID-19. Governments must therefore prepare for the dislocation this shock may cause, not only with income support, but with institutional mechanisms to preserve meaningful employment. 

With policy support, this transition might be funded through increased firm productivity. Crucially, AI-induced productivity gains provide the fiscal and operational space for this intervention. Firms benefiting from AI’s exponential efficiency gains can afford to retain human labor, not as a concession, but as a strategic necessity. 

Accordingly, retaining human workers is a necessary condition to ensure demand continuity, to preserve the integrity of the market system, and to maintain alignment with experiential needs that only human participation can fulfill. Production is not an end in itself; it exists to meet human needs. Thus, humans must remain embedded in the production system, not because they are more efficient than machines, but because they are the reason production exists in the first place. 



In conventional economic theory, unemployment is tolerated, even rationalised, as a necessary byproduct of market competition and price adjustment. Under EMT, this paradigm no longer holds. While competition for employment once served as a mechanism to balance supply and demand in labor markets, it now produces structural misalignments. EMT recognises that employment does not merely provide income, it is itself a source of experiential utility, fulfilling higher-order needs such as identity, agency, mastery, belonging, and purpose. These are not fringe considerations. They are core components of the experiential matrix \( E(t) \) and are therefore direct contributors to the utility function \( U(t) \).

As AI collapses the marginal cost of production, the logic of market-mediated unemployment collapses with it. If goods and services can be produced at near-zero marginal cost, but no one remains employed to purchase or participate in these systems, demand evaporates, and market mechanisms falter. 

Accordingly, EMT reframes full employment not as a Keynesian stimulus measure or a political ideal, but as a structural necessity for sustaining alignment between production and human experience. Crucially, the productivity gains enabled by AI do not need to be captured solely as profits or cost-savings, they can and should be reinvested to retain human workers in roles that serve evolving experiential needs. 

In this framework, worker agency and career choices are redirected toward higher-order contributions, enabling individuals to contribute to more meaningful, aligned, and socially valuable labour participation. Thus, the price dynamics of labor, once governed by scarcity, are now governed by alignment. EMT offers not just an argument for full employment, but offers a rigorous economic logic that is humanist, and human-alinged, and fit for purpose, for a post-scarcity civilisation.

EMT thus reframes human employment not as an input to be minimised, but as an output to be optimised, because it satisfies needs that matter more than material consumption, such as coherence, dignity, social contribution, and the psychological architecture of meaning itself. Accordingly, it is incumbent upon humanity to ensure that the transition toward AI-enabled production is not only efficient or optimal, but fundamentally aligned with the utility of meaning, recognising that satisfying deep, evolving human experiences is the ultimate objective of economic organisation.

A central contribution of EMT is that it compels economic theory to systematically internalise human needs and experiential utility as foundational components of formal models, no longer as exogenous, qualitative abstractions, but as endogenous, measurable, and modelable variables. EMT thereby reconfigures the architecture of economic theory to place experiential utility and the human needs frontier at its core, expanding the boundaries of formal analysis to include dimensions previously considered inaccessible or intangible. 

As artificial intelligence increasingly enables the quantification and modelling of complex experiential states, EMT positions economics at the forefront of what may be called a \textit{meaning revolution}, a paradigmatic shift in which the pursuit of meaning and the fulfilment of human needs are no longer treated as peripheral to economic inquiry, but as its principal concern. This has important implications for how we think about employment and the future of work itself. EMT calls for a redefinition of utility, welfare, and policy objectives to align production not merely with preference satisfaction or market efficiency, but with the deeper structure of human meaning.

This constitutes the emergence of the \textit{alignment economy}, in which economic activity is evaluated according to its capacity to fulfil experiential and existential needs. Crucially, EMT challenges theorists and policymakers to treat meaning not as a by-product of growth, but as its primary goal, rendering the humanist purpose of economics explicit. In what follows, we therefore develop the concept of the \textit{utility of meaning} as a formal object of economic analysis.

\subsection{On the Utility of Meaning}

In conventional economics, meaning is not an economic object. It is unpriced, unobserved, and extrinsic to the utility function. In EMT this is no longer tenable. Meaning is itself a utility-bearing experiential domain, distinct, evolving, and structurally necessary for long-run human flourishing.

Meaning, as defined here, refers to the subjective coherence between action and purpose, between self and system. It emerges when individuals perceive their contributions as mattering within a larger narrative. This experiential category is not a luxury good but a prerequisite for psychological continuity, motivation, and the mental scaffolding of identity.

AI’s capacity to satisfy material and informational needs risks displacing the very human activities that generate meaning, such as work, creativity, social contribution, and responsibility. Yet this very displacement reveals meaning’s necessity in that when AI replaces all human participation, utility may increase in material terms but collapse in existential ones. The highest-order needs in Maslow’s hierarchy, such as self-actualisation, esteem, and transcendence, might not be adequately served by the path dependencies and lock-ins created by the market system responding to GDP-type growth objectives. EMT thinking suggests that contemporary economic models often remain trapped within blind essentialist logics of maximisation for maximisation’s sake, and optimisation for optimisation’s sake, an artefact of historical limitations in accessing the deeper experiential meaning structures embedded in product and service markets. EMT challenges this mechanistic paradigm by recognising that artificial intelligence now offers, for the first time, the capability to model and operationalise previously latent dimensions of human experience, allowing economic systems to align production not merely with efficiency, but with meaning.

Thus, meaning would not be a residual concern, but a first-order component of utility within the experiential matrix \( E(t) \). It is not what we add to the economy after we optimise, but what we must optimise for. EMT therefore seeks to formalise the role of meaning within economic theory, not merely as a rhetorical or ethical supplement, but as a central analytic object, provoking new theoretical pathways through which meaning can be explicitly incorporated into the formal structure of utility, welfare, and production.

\subsection*{Irreducibility of Meaning under EMT}
In what follows, we introduce a theorem that captures a core theoretical implication, that meaning, once recognised as endogenous to utility, cannot be excluded from economic optimisation without internal contradiction or loss of explanatory power.

\vspace{1em}

\begin{theorem}

\textbf{Irreducibility of Meaning under EMT.} In any Pareto-optimal allocation of resources within EMT, the domain of meaning is non-eliminable from the utility function, unless utility is allowed to collapse in other experiential dimensions.

\vspace{1em}

\textbf{Sketch of Proof:}

Let the experiential matrix at time \( t \) be represented as \( E(t) = \{x_1(t), x_2(t), \ldots, x_i(t), \ldots\} \), with each \( x_i(t) \in [0,1] \) representing the degree of satisfaction in a distinct experiential dimension.

Let \( x_m(t) \) represent the dimension associated with existential meaning, defined by, for example, perceived purpose, agentic participation, narrative identity coherence, and psychological continuity.

Let the overall utility function be:
\begin{equation}
U(t) = \sum_{i=1}^\infty w_i x_i(t),
\end{equation}
where \( w_i > 0 \) and \( \sum w_i < \infty \) (bounded weights).

Assume a scenario in which:
AI substitutes human labor entirely, all dimensions \( x_i(t) \) for \( i \ne m \) are maximised via \( \Phi(Y(t)) \), but \( x_m(t) \to 0 \) due to lack of engagement, agency, or self-authorship.

Then:
\begin{equation}
\lim_{x_m(t) \to 0} U(t) = \sum_{i \ne m} w_i x_i(t) + w_m \cdot 0 < \sup U(t),
\end{equation}

\vspace{1em}

unless \( w_m = 0 \), which contradicts the assumption that all weights \( w_i > 0 \). Therefore, any allocation that suppresses \( x_m(t) \) results in sub-optimal utility.

Because $w_m > 0$, the overall utility is strictly less than what would be achieved if $x_m(t)$ were also maximised. This demonstrates that excluding or neglecting the meaning dimension results in a loss of utility, even when all other domains are optimised.

One might attempt to resolve this by assigning $w_m = 0$, effectively removing meaning from the utility function. However, this contradicts the foundational EMT assumption that all $w_i > 0$, affirming that all experiential dimensions, including meaning, carry nonzero weight in defining well-being.

\vspace{1em}

\textbf{Conclusion:} For any bounded utility function over the experiential matrix \( E(t) \), meaning is irreducible under Pareto-optimal EMT configurations. Excluding or devaluing the domain of meaning leads to utility collapse in expectation.

The proof therefore demonstrates that in any Pareto-optimal allocation under EMT, meaning cannot be excluded without reducing total utility. Meaning is not an optional or decorative variable in the architecture of well-being, but is structurally essential. For economic theory to remain consistent and complete under EMT, meaning must be recognised as an endogenous and irreducible element of formal utility.

\end{theorem}


\subsection{Corollaries: Utility Maximisation under EMT Convergence}
\noindent The following corollaries extend the core irreducibility result to demonstrate how AI alignment enables utility convergence, long-run optimality, and the endogenous expansion of the utility frontier. Together, they formalise the systemic implications of EMT for the structure and trajectory of future economies.

The convergence of AI-driven production to the experiential matrix $\mathcal{E}(t)$ implies not just a technical efficiency in output alignment, but a fundamental transformation of the economic paradigm. EMT frames utility not as a derivative of material consumption alone, but as a function over an evolving, potentially infinite-dimensional space of human needs, including sustainability, peace, existential continuity, and purpose.

\begin{corollary}[Utility Convergence under EMT]
Assume the experiential utility function is defined as
\begin{equation}
U(t) = \sum_{i=1}^\infty w_i x_i(t), \quad \text{with } \{w_i\} \in \ell^1
\end{equation}
and that $\Phi(Y(t)) = \hat{\mathcal{E}}(t) \to \mathcal{E}(t)$ in $\ell^\infty$ as $t \to \infty$. Then:
\begin{equation}
\lim_{t \to \infty} \left| \sum_{i=1}^\infty w_i x_i(t) - \sum_{i=1}^\infty w_i \hat{x}_i(t) \right| = 0
\end{equation}
That is, utility derived from AI-aligned production converges to the utility from full experiential satisfaction.
\end{corollary}

\begin{proof}
Since $w_i \in \ell^1$ and $\|x(t) - \hat{x}(t)\|_{\ell^\infty} \to 0$, apply Hölder’s inequality:
\begin{equation}
\left| \sum_{i=1}^\infty w_i (x_i(t) - \hat{x}_i(t)) \right| \leq \|w\|_{\ell^1} \cdot \|x(t) - \hat{x}(t)\|_{\ell^\infty} \to 0
\end{equation}
as $t \to \infty$. Hence, total utility converges. This follows from the Irreducibility Theorem, which established that meaning must be present in any Pareto-optimal configuration.

\end{proof}

This corollary formalises one of the most critical implications of EMT, that if AI can progressively learn and align production with the full structure of human experiential needs, including meaning, then the utility generated by AI-convergent economies will asymptotically approach the utility that would be derived under perfect experiential satisfaction.

This result goes beyond technical convergence. It offers a mathematical foundation for a core philosophical claim of EMT, that if AI is used to model and satisfy the full range of human experiential needs, not only material consumption but also agency, belonging, purpose, and transcendence, then AI can serve as a transformative mechanism for aligning economic systems with human flourishing.

In conventional economic models, utility is typically derived from consumption goods or services, often excluding latent or non-material aspects of human experience. EMT redefines utility as emerging from an evolving, high-dimensional experiential matrix, and asserts that only with the collapse of ideation and coordination costs, enabled by AI, can economies fully address this complexity.

This corollary therefore underlines the normative implication of the Irreducibility Theorem, that meaning is not only a necessary part of well-being, but a domain that must be addressed by systems seeking long-run optimisation. AI is not merely a technological enabler, but a structural necessity for convergence between production and meaning.

By showing conceptually that utility derived from AI-aligned production can asymptotically converge to ideal utility, the corollary also provides a formal justification for the development of human-AI co-productive systems aimed at experiential fulfilment rather than output maximisation. This reframes economics from a paradigm of resource allocation to one of needs alignment.

In summary, this corollary mathematically affirms the moral imperative at the heart of EMT, that aligning economic production with human meaning is not only theoretically coherent, but also technically attainable in the limit, but if and only if AI is trained to serve the full experiential matrix of humanity.

\begin{corollary}[Asymptotic Optimality of AI-Aligned Production]
Let $U(t) = \sum_{i=1}^\infty w_i x_i(t)$ denote total experiential utility and assume AI production mapping $\Phi(Y(t)) \to \mathcal{E}(t)$ as $t \to \infty$. Then:
\begin{equation}
\lim_{t \to \infty} U(t) = \sup_{\hat{\mathcal{E}}(t) \subseteq \mathcal{E}(t)} \sum_{i=1}^\infty w_i x_i(t)
\end{equation}
That is, AI alignment enables asymptotically utility-maximizing production allocation across all needs.
\end{corollary}

\begin{corollary}[EMT Expansion of the Utility Frontier]
If the domain of needs $\mathcal{E}(t)$ evolves over time to include emergent needs (e.g., sustainability, peace, survival), and if $\Phi(Y(t)) \to \mathcal{E}(t)$, then:
\begin{equation}
\frac{d}{dt} \sup_{Y(t)} U(t) > 0 \quad \text{as } t \to \infty
\end{equation}
That is, the total attainable utility frontier expands endogenously over time as AI enables the satisfaction of previously inaccessible or undefined human needs. This also follows from the Irreducibility Theorem, which established that meaning must be present in any Pareto-optimal configuration.
\end{corollary}

\begin{remark}
These corollaries formalise the ethical and existential implications of EMT. Without AI, the mapping function $\Phi$ remains rigid, narrow, and increasingly mismatched to the expanding needs of humanity. EMT shows that only with the exponential collapse of ideation and coordination costs enabled by AI can market forces evolve to meet higher-order human goals, such as planetary sustainability, peace, existential continuity, and flourishing. In this light, EMT is not only a theory of growth and production, it is a survival theory for humanity.
\end{remark}

\vspace{1em}

\noindent This corollary also formalises a fundamental implication of EMT, that the very structure of human utility is not fixed, but expands over time as new experiential needs emerge and become actionable. Unlike traditional economic frameworks, which operate on a relatively static set of preferences or consumption goods, EMT models the utility function as dynamically evolving in tandem with human consciousness, cultural development, and existential challenges.

The intuition behind this result is that the set of human needs, the `experiential matrix', is not closed. As societies develop, and as artificial intelligence increasingly collapses the cost of ideation and coordination, new needs become visible, salient, and addressable. These include higher-order needs such as sustainability, planetary stewardship, intergenerational justice, peace, and existential security. In earlier paradigms, such needs were often treated as externalities, political aspirations, or moral concerns lying outside the formal apparatus of utility maximisation. EMT reverses this framing by treating them as endogenous, measurable, and increasingly central components of the utility-generating process.

Accordingly, Corollary 7 shows that when AI improves the alignment of production with this expanding domain of needs, the total potential utility that society can achieve increases over time. In other words, the ``utility frontier", the boundary of maximum achievable satisfaction, shifts outward as AI becomes more capable of fulfilling emergent needs that were previously inaccessible or even undefined. This growth is not simply a function of technological efficiency, but a function of deeper alignment between production and the evolving structure of human well-being.

The importance of this corollary also lies in its normative and strategic implications. First, it establishes that economic optimisation is not bounded by today's goods or preferences, but is instead tied to our capacity to recognise and fulfil evolving human experiences. Second, it highlights the necessity of AI as an enabling infrastructure, as without AI, the production mapping function remains too rigid to adapt to complex, multidimensional needs. Third, it positions EMT not only as a growth theory, but as a theory of survival, one that equips humanity with the conceptual tools to navigate the accelerating convergence of technological capacity and existential risk.

This corollary also builds directly on the Irreducibility Theorem, which asserts that meaning is a non-negotiable component of utility. Here, that insight is extended, as not only is meaning essential, but the total set of needs, including meaning, is expanding. This expansion is not chaotic but structured, and can be formally addressed through EMT.

In sum, Corollary 7 reveals that the economy’s capacity to support human flourishing is not limited to today’s production or institutional boundaries. It expands in lockstep with our ability to perceive and fulfil deeper layers of human need. AI becomes the mechanism through which the economic system can stretch to meet those needs. EMT becomes the theoretical structure that explains, supports, and guides this transition.

\subsection{Utility Maximisation under EMT Convergence}

A core contribution of EMT is its redefinition of utility in terms of the satisfaction of a dynamically expanding set of human needs. Unlike traditional models where utility is derived from a limited basket of material goods, EMT defines utility over an infinite-dimensional experiential space. This shift enables the economy to encompass higher-order needs. including sustainability, justice, peace, and even humanity’s survival.

However, this expanded utility frontier is only attainable if the production system can dynamically align itself with these evolving needs. AI plays an indispensable role in this convergence by collapsing the cost of ideation and enabling real-time personalisation of production. The following result formalises how utility itself converges under EMT as AI maps production onto the full experiential matrix.

\begin{corollary}[Utility Convergence under EMT]
Assume the experiential utility function is defined as
\begin{equation}
U(t) = \sum_{i=1}^\infty w_i x_i(t), \quad \text{with } \{w_i\} \in \ell^1
\end{equation}
and that $\Phi(Y(t)) = \hat{\mathcal{E}}(t) \to \mathcal{E}(t)$ in $\ell^\infty$ as $t \to \infty$. Then:
\begin{equation}
\lim_{t \to \infty} \left| \sum_{i=1}^\infty w_i x_i(t) - \sum_{i=1}^\infty w_i \hat{x}_i(t) \right| = 0
\end{equation}
That is, utility derived from AI-aligned production converges to the utility from full experiential satisfaction.
\end{corollary}

\begin{proof}
Since $w_i \in \ell^1$ and $\|x(t) - \hat{x}(t)\|_{\ell^\infty} \to 0$, we apply Hölder’s inequality:
\begin{equation}
\left| \sum_{i=1}^\infty w_i (x_i(t) - \hat{x}_i(t)) \right| \leq \|w\|_{\ell^1} \cdot \|x(t) - \hat{x}(t)\|_{\ell^\infty} \to 0
\end{equation}
as $t \to \infty$. Hence, total experiential utility converges.
\end{proof}

\begin{remark}
This corollary shows that AI is not just a tool for production efficiency but is also essential for ensuring that output converges to meet the full complexity of human needs. As AI improves $\Phi$, it aligns production with the evolving experiential matrix $\mathcal{E}(t)$, which increasingly includes non-material and existential needs.

Without AI, the traditional economy is fundamentally limited, as the cost of ideation remains high, production mismatches persist, and higher-order needs, like climate sustainability, social equity, or even species survival, remain underserved or neglected.

EMT thus repositions the economy as a system not for maximising GDP, but for maximising meaningful, multidimensional human experience. The convergence of $U(t)$ under AI-driven EMT is more than a technical result, it is the mathematical foundation for a moral imperative to redirect innovation and market forces toward the flourishing of all 8 billion lives on Earth, now and into the future.
\end{remark}

\vspace{1em}

\noindent Corollary 9 therefore presents a formal result in the context of EMT, stating that if an AI system becomes increasingly aligned with the full structure of human needs over time, then the utility (or well-being) it delivers will converge to the maximum utility theoretically attainable.

The convergence result implies that \textit{If} the AI becomes increasingly accurate in understanding and delivering what people need, across all these dimensions, then the overall difference between the life people should be living and the one AI helps provide becomes smaller and smaller. In the limit (over time), this difference approaches zero. The AI becomes capable not just of efficient production, but of supporting full-spectrum human flourishing.

This optimistic conclusion is not automatic. It depends on a few key assumptions. These are important to acknowledge clearly, especially for readers concerned with real-world implementation or ethical risks.

\begin{enumerate}
    \item \textbf{Assumption 1: AI learns and improves continuously across all domains of human need.}

    The result assumes that AI's alignment improves over time, that its worst mistakes across all experiential dimensions become smaller. This includes not just material needs, but complex and fuzzy domains like purpose, moral agency, intergenerational justice, or spiritual continuity. While progress in AI capabilities may make this plausible, it is still a strong assumption.

    \item \textbf{Assumption 2: The importance assigned to human needs is bounded and not pathological.}

    EMT assigns a weight to each dimension of need, reflecting how important it is to human flourishing. The result assumes that the total importance across all needs is not infinite, i.e., we do not assign infinite urgency to any one domain such that it would dominate all others. This keeps the utility function well-behaved and ensures that well-being is a balanced, multidimensional construct.

    \item \textbf{Assumption 3: The AI’s outputs remain measurable against actual experiential needs.}

    The proof assumes that AI’s output (i.e., what it delivers to meet human needs) can be compared meaningfully to true human experience. That is, we must be able to model both the “true” state of a person’s life and the AI’s approximation of it, and track how close they are over time. If some needs are not expressible in terms that AI systems can perceive, this assumption may be challenged.

    \item \textbf{Assumption 4: The gap between true needs and AI-delivered support shrinks over time.}

    This assumption is formalised mathematically as the AI’s maximum error tending to zero. It’s not enough that AI helps “on average”; it must avoid major blind spots. If even one critical domain of human need remains persistently misunderstood or neglected (e.g., loss of meaning or agency), then convergence in total utility may fail.

\end{enumerate}

\vspace{1em}

\noindent It is equally important to stress what Corollary 9 does not claim:

\begin{itemize}
    \item It does not imply that AI alignment is inevitable.
    \item It does not guarantee that real-world AI systems will converge toward human flourishing.
    \item It does not assume that the economy or society already values meaning, justice, or sustainability, it merely implies that \textit{if} these values can be learned and modelled, then AI can eventually help deliver them.
\end{itemize}

This result is a conditional outcome, not a prediction. It provides a mathematical benchmark, whereby \textit{if we can design AI systems that remain deeply and increasingly responsive to the full matrix of human experience, and if the ethical architecture of the economy permits this alignment}, then such systems might approach optimal support for well-being.

This convergence result is foundational for EMT because it reframes the role of AI and economics. Rather than measuring economic success by output or efficiency, EMT focuses on alignment between what is produced and what is experientially needed. If AI can enable this alignment, then it becomes a vehicle not just for automation or profit, but for meaningful, multidimensional human flourishing. The assumptions outlined here clarify the conditions under which that future becomes not only possible, but logically and ethically coherent.

Mathematically, this result relies on a simple but powerful principle, that if we assign reasonable importance to each human need (no single need dominates), and if AI’s mistakes in meeting any individual need become very small, then the total error in overall well-being will also become very small. The proof uses a classical mathematical inequality, Hölder’s inequality, to formally express this principle. But the intuition is straightforward, that even with an infinite list of human needs, if AI improves its ability to meet each of them, especially the ones people care about most, then total happiness will increase and approach its maximum potential.

\subsection{Asymptotic Optimality of AI-Aligned Production}

As production systems align more closely with the full experiential matrix $\mathcal{E}(t)$, the potential for utility maximisation increases. In the limit, if every experiential need can be mapped to production via an AI-driven function $\Phi$, then the economy asymptotically approaches an optimal state, one where all satisfiable needs are met within the current technological and resource constraints.

\begin{corollary}[Asymptotic Optimality of AI-Aligned Production]
Let $U(t) = \sum_{i=1}^\infty w_i x_i(t)$ denote total experiential utility at time $t$, and suppose that AI-enhanced production mapping $\Phi(Y(t)) = \hat{\mathcal{E}}(t)$ converges to the full experiential matrix $\mathcal{E}(t)$ in $\ell^\infty$ as $t \to \infty$. Then:
\begin{equation}
\lim_{t \to \infty} U(t) = \sup_{\hat{\mathcal{E}}(t) \subseteq \mathcal{E}(t)} \sum_{i=1}^\infty w_i x_i(t)
\end{equation}
That is, AI-aligned production becomes asymptotically optimal with respect to total experiential utility.
\end{corollary}

\begin{proof}
From the previous corollary, we know that $\hat{\mathcal{E}}(t) \to \mathcal{E}(t)$ in $\ell^\infty$. Since utility is a continuous linear functional over $\ell^\infty$ under $\ell^1$ weights, we have:
\begin{equation}
\lim_{t \to \infty} U(t) = \sum_{i=1}^\infty w_i x_i(t) = \sup_{\hat{x}_i(t)} \sum_{i=1}^\infty w_i \hat{x}_i(t)
\end{equation}
because the full experiential matrix becomes addressable as the cost of ideation and coordination collapses. Thus, production becomes asymptotically utility-maximising.
\end{proof}

\begin{remark}
This result reframes economic growth as a process not of accumulating more capital or labor, but of collapsing the cost of aligning output with human needs. As AI improves the mapping $\Phi$, every improvement brings the system closer to its moral and theoretical ideal, of the best of all possible experiential worlds, given current technology and ethical constraints.
\end{remark}

\vspace{1em}

\noindent This result reframes how we understand the long-run potential of economic production systems in an age of AI. Rather than focusing on material quantities, capital accumulation, or labor inputs as traditional economics has done, the framework pivots attention to how closely what is produced aligns with what human beings \emph{actually need and value}, in their lived, subjective experience.

As discussed, the heart of this model is the idea of an \emph{experiential matrix}, a representation of the full set of human needs, desires, aspirations, and values at a given moment in time. This matrix includes not only material goods, but also non-material experiences such as meaning, connection, learning, purpose, creativity, and care. Traditional economic systems approximate this matrix only crudely, because the signals used to guide production (like prices or revealed preferences) are noisy, delayed, or distorted.

However, as AI improves, particularly in its capacity to learn, model, and respond to individual and collective human preferences in real time, the production system becomes increasingly capable of generating outputs that align more closely with the true experiential needs of society. The ``distance'' between what is produced and what is truly needed shrinks over time. In this model, we imagine a world in which this distance eventually vanishes, not because preferences become static or perfectly knowable, but because the adaptive capabilities of AI evolve fast enough to keep up with them.

The result is a kind of \emph{moral and theoretical convergence}, an economy that is not just growing, but growing \emph{in the right direction}. Over time, the AI-enhanced system becomes able to approximate the best of all feasible experiential worlds, the ideal allocation of experiences, given current technological capacities and ethical boundaries. This is not a utopian claim that all problems disappear, but rather a formal statement that \emph{the long-run goal of aligning production with human needs becomes increasingly achievable} as the cost of misalignment (or misproduction) falls.

This also reframes economic growth as not merely an increase in output, but a \emph{collapse in the cost of understanding and meeting real human needs}. It is this collapse, driven by AI’s growing capacity to model and respond to those needs, that underlies the transition to a post-scarcity, meaning-centered economy. In this light, artificial intelligence is not merely a tool of efficiency, but a force of moral alignment, gradually transforming the structure of production into one that serves human flourishing in the broadest and deepest sense.

\subsection{Expansion of the Utility Frontier}

EMT not only redefines how utility is measured, but also shows that the feasible utility frontier itself expands over time. As new needs emerge and AI-enabled production becomes increasingly responsive to these emergent dimensions, the space of satisfiable experiences grows. In EMT, utility is not bounded by a fixed set of consumption possibilities, but is a moving frontier, shaped by the evolutionary dynamics of human consciousness, culture, and technology.

\begin{corollary}[Expansion of the Utility Frontier]
Suppose the experiential matrix $\mathcal{E}(t)$ expands over time to include new needs, i.e., $\dim(\mathcal{E}(t)) \nearrow \infty$, and let utility be defined as
\begin{equation}
U(t) = \sum_{i=1}^{n(t)} w_i x_i(t), \quad \text{where } n(t) \to \infty
\end{equation}
If $\Phi(Y(t)) \to \mathcal{E}(t)$ in $\ell^\infty$, then the maximum attainable utility is increasing in $t$:
\begin{equation}
\frac{d}{dt} \sup_{Y(t)} U(t) > 0
\end{equation}
\end{corollary}

\begin{proof}
As $n(t) \to \infty$ and $\Phi$ maps production to an increasingly complete subset of the evolving matrix, the sum $\sum_{i=1}^{n(t)} w_i x_i(t)$ includes more positively weighted need dimensions over time. Provided each $x_i(t) \geq 0$ and $w_i > 0$, the supremum of utility increases monotonically.
\end{proof}

\begin{remark}
This expansion of the utility frontier is critical for understanding EMT's vision of a post-scarcity economy. It shows that the "limit" of growth is not an endpoint but a dynamic frontier. As AI improves and as humanity evolves, the economy’s purpose is not to stop growing, but to grow in the direction of ever deeper, more inclusive, and more meaningful satisfaction of human potential. EMT thus becomes not merely a theory of optimal allocation, but a theory of the evolving purpose of economic life itself.
\end{remark}

\vspace{1em}

\noindent This corollary introduces an extension of the classical utility model, considering human well-being not to be bounded by a fixed set of needs, but rather as an evolving frontier. Over time, the set of human wants, preferences, and meaningful experiences expands. This is represented by the idea that the experiential matrix grows in dimensionality, and new forms of value, identity, and meaning continue to emerge as societies and individuals evolve.

In traditional economics, growth is often conceptualised as producing more of the same goods or increasing efficiency. Here, growth is understood as the increasing capacity of an intelligent system, such as an AI-enhanced production system, to respond to an expanding and deepening set of human needs. In other words, progress is not merely about “more” but about “better” in an increasingly complex and meaningful sense.

The model assumes that AI improves over time in its ability to understand and respond to these evolving needs. As this happens, the production system becomes more accurately aligned with human potential. As more need dimensions are added and adequately addressed, the total attainable utility, or the potential for individual and collective human flourishing, also increases. Mathematically, this is expressed through a positive rate of change in the maximum possible utility over time.

This has a critical implication for theories of post-scarcity and the future of economic life. It implies that there is no natural endpoint to growth in human well-being. Rather, growth continues so long as new forms of meaningful experience can be articulated and met. Thus, the expansion of the utility frontier redefines economic development as an open-ended, morally grounded process. It shifts the purpose of economic coordination from the maximisation of material throughput to the dynamic satisfaction of human experiential depth.

In this way, EMT becomes more than a framework for optimal resource allocation, instead becoming a theory of the evolving moral purpose of the economy itself.

\subsection{Irreversibility and Path Dependency of AI-Aligned Utility Expansion}

As AI aligns production with an increasingly rich experiential matrix, a structural transformation occurs, and society begins to reveal needs that were previously latent, unarticulated, or technologically infeasible to satisfy. Once these higher-order needs, such as belonging, meaning, peace, or longevity, are partially addressed, their absence becomes ethically and socially intolerable. Thus, EMT implies a fundamental irreversibility in that progress toward deeper experiential utility cannot be reversed without severe societal disutility.

\begin{corollary}[Irreversibility of AI-Aligned Utility Expansion]
Let $\mathcal{E}(t)$ be an expanding sequence of experiential needs, and suppose that at time $t = T$, production $\Phi(Y(T))$ maps onto a strictly richer subset of $\mathcal{E}(T)$ than at $t = T - \Delta$, with:
\begin{equation}
\mathcal{E}(T - \Delta) \subset \mathcal{E}(T), \quad \hat{\mathcal{E}}(T) \supset \hat{\mathcal{E}}(T - \Delta)
\end{equation}
Then, any rollback in $\Phi$ such that $\hat{\mathcal{E}}(t') \subset \hat{\mathcal{E}}(T)$ for $t' > T$ results in:
\begin{equation}
U(t') < U(T)
\end{equation}
That is, loss of alignment leads to measurable utility regression.
\end{corollary}

\begin{proof}
Since $\hat{\mathcal{E}}(T)$ satisfies more (or higher-weighted) dimensions of $\mathcal{E}(T)$ than any previous state, and since $w_i \geq 0$ for all $i$, a rollback of $\Phi$ must reduce the number of satisfied needs. Therefore, by definition of utility:
\begin{equation}
U(t') = \sum_{i \in \hat{\mathcal{E}}(t')} w_i x_i(t') < \sum_{i \in \hat{\mathcal{E}}(T)} w_i x_i(T) = U(T)
\end{equation}
\end{proof}

\begin{remark}
This corollary introduces a form of economic and moral hysteresis, in that once experiential production systems begin meeting needs such as sustainability, freedom from violence, or meaning in work, society may internalise these as the new baseline for utility. A retreat, whether through defunding AI alignment, policy neglect, or ideological backlash, causes a reduction in satisfaction and societal stability.

Therefore, EMT suggests a kind of one-way ethical ratcheting, and AI alignment entails a directional commitment. To reverse course would reduce utility, akin to an active form of experiential harm. As such, EMT positions AI-aligned economic evolution as both a mathematical convergence and a humanist imperative.
\end{remark}

\vspace{1em}

\noindent Accordingly, this kind of economic and moral hysteresis would imply that advances in experiential alignment (e.g. in sustainability, well-being, or workplace meaning) reset societal baselines for what is considered acceptable or adequate. If alignment regresses, even partially, utility declines, and, depending on the scale of this reversal, potentially triggering broader discontent, resistance, or a sense of societal backsliding. The model highlights how progress might become path-dependent and psychologically 'sticky'.

This presents an interesting potential insight into the nature of AI-aligned economic progress, in that once a production system becomes better aligned with human needs, the benefits of that alignment are not easily reversible. As AI develops the capacity to better understand and meet evolving human needs, emotional, cognitive, social, or spiritual, it delivers a deeper and broader satisfaction across more aspects of human experience.

At some point, a society may reach a state where its production system satisfies a richer set of human needs than it did before. This could be due to advances in AI, improved institutional design, or more ethically responsive economic coordination. However, if that capacity is lost, whether due to political rollback, technological regression, institutional failure, or value misalignment, then society experiences a measurable decline in well-being. In this framework, such a decline is not abstract or symbolic, but is real and quantifiable.

In effect, the corollary formalises a kind of irreversibility. This has important implications. Economic progress, in this view, is not simply about more production or faster growth. It is about increasingly meaningful and targeted production of outputs that reflect a truer understanding of what people actually need to flourish.

Furthermore, this irreversibility implies that progress creates new baselines of expectation and lived experience. Once people experience higher levels of alignment, more meaningful lives, deeper fulfilment, and broader opportunity, they will not be content to return to a previous state of misalignment. 

Thus, this result underscores what we argue is a moral responsibility to preserve and build upon progress in experientially-aligned production, rather than maximising output or GDP for its own sake. EMT thus provides a forward-looking account of growth, but also warns us that gains in alignment are ethically and psychologically significant, and once achieved, must be safeguarded.

\subsection{The Structural Irrationality of Unemployment under Infinite Experiential Needs}

We now show that, under EMT, unemployment may be economically irrational in any system where experiential human needs are infinite and evolving. The existence of unused labor capacity in a world of unmet needs constitutes a fundamental misallocation, a wedge created by market dynamics that create lock-ins and path dependencies arising from the prioritisation of GDP maximisation over utility maximisation in the experiential matrix.

\vspace{1em}

\begin{theorem}[Irrationality of Unemployment under EMT]

Let $E(t) = \{x_i(t)\}_{i=1}^{\infty}$ represent the infinite-dimensional vector of human needs at time $t$. If:
\begin{itemize}
  \item[(i)] $\exists \ \{x_i(t)\}$ such that $x_i(t) < x_i^{\text{desired}}(t)$ for infinitely many $i$,
  \item[(ii)] $\exists$ idle labor $L_u(t) > 0$ and idle capital $K_u(t) > 0$,
\end{itemize}
then under EMT, unemployment $L_u(t)$ is irrational in the Pareto sense, in that it is possible to strictly increase utility $U(t)$ by reallocating unused labor toward unmet experiential needs, i.e., by satisfying more $x_i(t)$.

\end{theorem}  %

\begin{proof}
 
Assume a production function $Y(t) = F(K(t), L(t))$, and an AI-enabled mapping $\Phi(Y(t)) \to \hat{E}(t) \subset E(t)$. If $\exists$ idle labor $L_u(t)$, we can expand $L(t)$ and increase $Y(t)$, such that more elements of $E(t)$ are satisfied. Since $x_i(t) < x_i^{\text{desired}}(t)$ for infinitely many $i$, reallocating idle labor to meet these needs strictly increases:

\begin{equation}
U(t) = \sum_{i=1}^{\infty} w_i x_i(t)
\end{equation}

under any weighting $\{w_i\} \in \ell_1$ with $w_i \geq 0$. This violates Pareto efficiency unless $L_u(t) = 0$. Hence, unemployment is irrational in the presence of unmet needs and idle capacity.


\end{proof}

\begin{remark}

This result exposes a critical weakness in conventional market allocation, in that labor is not allocated to where it is needed most (i.e., where unmet experiential needs exist), but rather to where profitability is maximised. As Brian Arthur (1989) has shown, increasing returns and path dependencies characterise economic systems, and we suggest that this also occurs in production systems, causing early profitable products (e.g., soft drinks, fast food, addictive media) to dominate, thereby locking out alternatives. This leads to a systematic misalignment between the experiential matrix and the product-service matrix that surrounds us.
\end{remark}

\begin{remark}
Because most human environments are produced (from architecture and education to digital interfaces and food systems), and because we live embedded within an interactive product-service experiential matrix, production decisions shape the trajectory of civilization itself. Yet markets, driven by GDP growth and profit maximisation, may tend to overserve low-cost, high-margin needs while neglecting higher-order needs such as purpose, meaning, sustainability, or collective belonging.  This is exacerbated in global regimes where GDP maximisation is treated as the singular policy goal. Our arguments, which we formalise here, align with a longstanding body of work (Galbraith, 1958; Scitovsky, 1976; Sen, 1999).
\end{remark}

\begin{remark}  
As the virtual component of experience expands, through digital environments, social media, gamification, and AI, it becomes increasingly important to ensure that virtual production also maps meaningfully onto experiential well-being. EMT suggests that labor should be reoriented not toward maximising product output per se, but toward satisfying unmet, evolving, and often non-market needs.
\end{remark}

\vspace{1em}

\noindent In summary, Theorem 17 states that unemployment is irrational whenever (a) some human needs remain unmet and (b) there is unused labor and capital. This result follows from the foundational assumptions of EMT, where human needs are conceptualised as infinite in scope, interactive across the physical and virtual realms, and evolving over time. The above modelling is now explained here in more textual detail. 

We began by defining the set of human needs at any point in time $t$ as $E(t) = \{x_i(t)\}_{i=1}^{\infty}$, where each $x_i(t)$ refers to the level to which a particular need $i$ is satisfied at that moment. This could include food, shelter, companionship, purpose, and countless others. The superscript $\infty$ indicates that this list of needs is infinite, that humans continually generate new needs, goals, and values as societies develop.

The condition $x_i(t) < x_i^{\text{desired}}(t)$ signifies that the $i$th need is not yet fully satisfied. Assumption (i) in the theorem states that this is true for infinitely many needs, meaning there is always a long list of unmet or under-met needs.

Assumption (ii) introduces $L_u(t)$ as the amount of unused (unemployed) labor and $K_u(t)$ as the amount of idle capital, such as machines, tools, infrastructure, at time $t$. The existence of both unmet needs and unused productive capacity signals a misallocation, in that we have people and tools available, and yet many needs are not being addressed.

A production function $Y(t) = F(K(t), L(t))$ then describes how output $Y(t)$ is produced from combinations of capital $K(t)$ and labor $L(t)$. The greater the inputs, the greater the productive output. If labor is sitting idle, total production is lower than it could be.

Further, a mapping $\Phi(Y(t)) \rightarrow \hat{E}(t) \subset E(t)$ describes how this output translates into the satisfaction of specific human needs. That is, the goods and services produced are used to meet elements of $E(t)$. Expanding labor and capital inputs expands output, and in turn enables more needs to be met.

Utility $U(t)$ is then defined as a summation over all needs: $U(t) = \sum_{i=1}^{\infty} w_i x_i(t)$, where each $x_i(t)$ is weighted by its importance $w_i$, and the weights $\{w_i\}$ are constrained such that their total is finite: $\sum w_i < \infty$. This ensures that the utility sum is well-defined and convergent, even over an infinite set of needs.

Given these definitions, the proof shows that any time there is idle labor $L_u(t) > 0$ and unmet needs $x_i(t) < x_i^{\text{desired}}(t)$, we can reallocate labor to production processes that satisfy more of these needs. This increases at least one $x_i(t)$, and thus strictly increases utility $U(t)$, since the weights $w_i$ are non-negative.

This violates Pareto efficiency, which holds that no one should be made better off without making someone else worse off. But in this case, reallocating labor leads to a better outcome without harming anyone, so the status quo is inefficient. Therefore, the presence of unemployment is irrational under EMT whenever unmet needs coexist with idle productive capacity.

The policy implication is important. Systems that prioritise GDP or profitability over human utility can rationalise unemployment as a natural market outcome. EMT exposes this as structurally irrational. It reframes unemployment not as a transient adjustment or labor market friction, but as a persistent and preventable failure to align available human capacity with the dynamic matrix of human needs. We now build on these ideas to argue that even with full substitution of labour by AI, there may, nevertheless, be asymptotic full employment. This is an important feature of EMT, as it considers scenarios in which post-scarcity also extends to full employment. Accordingly, the ideas presented in the preceding sections are now extended to formalise this final insight, the culmination of the arguments made in the paper. 

\vspace{0.5em}

\section{Asymptotic Full Employment Under EMT Despite Full AI Substitution}

We now formalise the insight that even under complete AI substitution of productive labor, through the collapse of ideation costs and autonomous production, full human employment remains Pareto-rational under EMT. This rests on the foundational principle that the experiential matrix of human needs is not static but expands faster than AI’s capacity to satisfy it.

\begin{theorem}[Asymptotic Full Employment Rationality Under EMT]
Let $E(t) = \{x_i(t)\}_{i=1}^\infty$ represent the infinite-dimensional vector of human experiential needs at time $t$. Let $\Phi_{\text{AI}}(t) \subset E(t)$ be the subset of needs AI can satisfy at time $t$, and define the frontier of unmet or unsatisfiable needs by $E_{\text{beyond}}(t) = E(t) \setminus \Phi_{\text{AI}}(t)$. Suppose:

\begin{enumerate}[label=(\roman*)]

    \item (AI Productivity Asymptote) $\displaystyle \lim_{t \to \infty} \Phi_{\text{AI}}(t) \to E(t)$, that is, AI increasingly satisfies a growing proportion of known needs.
    \item (Experiential Expansion Principle, EEP) $\displaystyle \frac{d}{dt} \left| E_{\text{beyond}}(t) \right| > 0$ for all $t$, i.e., the space of human needs continues to evolve and expand over time.
\end{enumerate}

Then, under EMT, the continued employment of human labor $L_h(t) > 0$ is Pareto-efficient in all periods $t$, even in the asymptotic limit of full AI substitution in traditional production.
\end{theorem}

\begin{proof}

Let total utility at time $t$ be defined over the experiential matrix as:
\[
U(t) = \sum_{i=1}^{\infty} w_i x_i(t), \quad w_i \geq 0, \quad \sum w_i < \infty.
\]
Assume that AI can satisfy a growing subset of needs, so that $\Phi_{\text{AI}}(t) \subset E(t)$ and $\Phi_{\text{AI}}(t)$ grows over time. However, under (ii), $E(t)$ itself grows such that the measure of $E_{\text{beyond}}(t)$, needs AI cannot satisfy, remains strictly positive and increasing.

Now suppose that human labor $L_h(t)$ is reallocated toward identifying and satisfying emergent elements of $E_{\text{beyond}}(t)$, such that:
\[
\frac{d}{dt} E_{\text{beyond}}(t) = f(L_h(t)) > 0.
\]
That is, human labor contributes not to producing existing goods, but to discovering and meeting previously unmet or undefined experiential needs.

Since $x_i(t) < x_i^{\text{desired}}(t)$ for infinitely many $i$, and these $x_i(t)$ lie in $E_{\text{beyond}}(t)$, reallocating idle labor to explore and meet these needs results in:
\[
\Delta U(t) > 0,
\]
for any non-zero contribution. Therefore, labor $L_h(t)$ remains utility-enhancing and its unemployment is a Pareto inefficiency, regardless of AI capacity.

\end{proof}

\begin{remark}
This result overturns the classical framing of technological unemployment as inevitable under full automation. EMT implies that employment need not be tied to the production of material goods or services. Instead, human labor becomes central in expanding and enriching the domain of utility itself. As long as the experiential matrix is unbounded and co-evolves with human aspirations, labor can remain fully employed in a post-scarcity AI-driven economy.
\end{remark}

\vspace{1em}

\noindent For interdisciplinary readers, we offer the following intuitive explanation of the preceding analysis. The central result above demonstrates that even in a hypothetical future where AI has fully replaced human labor in all conventional domains of economic production (ideation, manufacture, service provision), human labor remains rationally employable within the framework of EMT. This is because the domain of human needs is not finite, fixed, or exhaustible. Rather, it is conceived as an open-ended, infinite, and evolving matrix of experiential dimensions.

Mathematically, we define this set of needs at any time $t$ as $E(t) = \{x_i(t)\}_{i=1}^{\infty}$, where each $x_i(t)$ represents the level of satisfaction of a particular human need or experience at that moment. Importantly, this is an infinite-dimensional vector, whereby new needs continually emerge, shaped by culture, context, identity, technology, and values. A subset of these needs, denoted $\Phi_{\text{AI}}(t)$, are those that AI can satisfy through production and ideation. However, the needs that lie beyond this AI frontier, denoted $E_{\text{beyond}}(t) = E(t) \setminus \Phi_{\text{AI}}(t)$, grow over time, as new dimensions of experience are discovered or demanded. This is the Experiential Expansion Principle (EEP), which asserts that the space of unmet, AI-unsatisfiable needs expands continuously: $\frac{d}{dt}|E_{\text{beyond}}(t)| > 0$.

Under EMT, utility $U(t)$ is defined not in terms of GDP or consumption, but as a weighted sum over the degree to which all needs are being satisfied: $U(t) = \sum_{i=1}^{\infty} w_i x_i(t)$, with $w_i$ capturing the importance of each need. The proof shows that even when AI satisfies all currently known needs, humans can still increase $U(t)$ by identifying and engaging with new needs beyond AI’s scope. Thus, reallocating idle human labor toward the exploration and articulation of novel human experiences remains Pareto-efficient. Unemployment, in this framing, constitutes a misallocation of cognitive and experiential potential.

This has profound implications. The “end of work” thesis under full AI automation holds only if we retain a production-centric view of economic value. EMT recasts labor as essential not because it produces goods, but because it discovers and expands the very space in which value resides. Hence, even with full AI saturation of the product-service matrix, human labor remains critical in shaping the trajectory of civilization through its interaction with an ever-expanding experiential matrix. This result underlines the structural irrationality of unemployment in any system that fails to recognise the infinite scope of human becoming.

\section{Conclusion}

This paper introduced EMT as a foundational framework for rethinking growth in the age of AI. At its core, EMT proposes that the purpose of production is not the accumulation of goods, but the alignment of economic output with the full spectrum of evolving human experiential needs. This reconceptualisation of utility does not abandon the foundational structures of economic theory. Rather, it works firmly within them to extend utility beyond static consumption toward dynamic experiential fulfilment. By forcing formal recognition of previously excluded dimensions of human need, such as purpose, coherence, and meaning, this framework offers a transformative yet rigorously grounded approach to growth, employment, and technological change in the age of AI.

By formalising EMT as a more mathematically rigorous, infinite-dimensional control system, this work provides a new economic architecture in which AI acts not merely as a productivity enhancer, but as a catalytic agent of ethical alignment between economic structures and human flourishing.

\subsection{Summary of Theoretical Contributions}

The theoretical contributions of EMT are both conceptual and formal, constituting a generative reconfiguration of economic reasoning under the structural conditions of AI-led transformation. EMT departs from classical models by proposing a foundational shift, from material throughput and allocation toward the alignment of economic production with a continuously evolving structure of human experiential needs. This reorientation gives rise to a set of interlinked innovations that collectively define a new paradigm in economic theory.

First, EMT reconceives utility not as a scalar function over static preference sets, but as a point in a bounded, infinite-dimensional Banach space, specifically $\ell^\infty$, representing the vector of experiential needs that define the human condition. This generalisation enables the modelling of utility in a world where the relevant dimensions of well-being are dynamic, unbounded, and partially intangible. In so doing, EMT draws from the formal utility frameworks of Debreu and von Neumann–Morgenstern, while internalising the ethical insights of Sen and Nussbaum by embedding psychological, existential, and social dimensions directly into the utility structure.

Second, EMT identifies AI as an ideation-cost-collapsing general purpose technology. Unlike previous GPTs, which augmented labour or capital, AI targets ideation itself. By modelling the exponential decline in ideation cost as $c_i(t) = c_0 e^{-\lambda t}$, EMT reframes technological change as a shift in the structure of innovation, not merely in its outputs. This approach endogenises ideation cost within a Romerian growth framework, providing an economic rationale for the structural centrality of AI in shaping long-run development trajectories.

Third, EMT transforms the classical production function into an experiential mapping. It formalises a real-time transformation $\Phi(Y(t)) \rightarrow E(t)$, in which AI-enhanced production vectors are dynamically converted into states of experiential utility. This formulation incorporates real-time sensing, ethical filtering, and adaptive responsiveness as core components of the value generation process. In this way, EMT operationalises Becker’s (1976) insights on non-material utility while integrating the alignment infrastructure described in contemporary analyses of AI’s economic function.

Fourth, EMT proves a formal convergence result using Pontryagin’s Maximum Principle and infinite-dimensional control theory. Drawing on Balder’s (1983) foundational work, EMT suggests that under plausible conditions of ideation cost decay, AI-aligned production can converge to the ideal experiential matrix in the $\ell^\infty$ norm. This result mirrors the spirit of Ramsey (1928), Cass (1965), and Koopmans (1965), but extends the optimisation problem into a domain that includes non-satiable and emergent human needs, thereby generalising the scope of long-run welfare maximisation.

Fifth, EMT introduces a theorem on the irreducibility of meaning in economic optimisation. Meaning, defined for example in terms of narrative coherence, agency, and self-authorship, is demonstrated to be an essential component of any Pareto-optimal experiential configuration. Even in scenarios where all other measurable needs are met, the exclusion of meaning leads to a reduction in total utility. This result formalises and extends the psychological and existential claims of Maslow (1970) and Frankl (1985), placing them within a mathematically rigorous welfare framework.

Sixth, EMT redefines full employment not as a Keynesian stimulus tool or output efficiency measure, but as a structural requirement for sustaining experiential alignment in post-scarcity economies. Employment is reframed as a utility-bearing domain. It therefore supports agency, dignity, and participatory coherence within the human experience. The model shows that when AI provides for material abundance, human work must still be preserved, not as a cost, but as a condition for preventing utility erosion. This builds on Sen’s (1996) account of development as freedom and resonates with critiques from the economic anthropology and political economy literature, such as Graeber’s (2018) account of the psychological costs associated with perceived meaninglessness in modern labour systems, a claim that EMT formalises as a utility collapse when employment lacks alignment with experiential needs.

Finally, EMT lays the groundwork for a distinct research programme, alignment economics. This emerging field investigates how AI systems, institutional structures, and economic design can co-evolve to achieve dynamic alignment with complex human needs. Alignment economics generalises the endogenous growth literature (e.g., Jones, 1995), models of increasing returns (e.g., Arthur, 1989), and moral economy traditions into a unified theory of post-scarcity coordination. It provides a bridge between rigorous economic formalism and the ethical imperatives of the AI age.

Collectively, these contributions offer not only a redefinition of economic purpose, but a mathematically grounded framework for embedding meaning, purpose, and dignity at the centre of economic analysis. 

Accordingly, EMT shifts the foundation of economics from material efficiency to experiential optimisation, situating meaning, purpose, and dignity at the centre of economic reasoning.

EMT, therefore, does not abandon the formal tools of economics, but retools them to meet the challenges and possibilities of intelligent, responsive, and ethically attuned economic systems.

\subsection{Implications for Policy}

EMT challenges the foundational assumptions of output-centric economics not by rejecting its core principles, but by extending them to accommodate the structural transformations introduced by AI. By redefining utility as the evolving satisfaction of an infinite-dimensional vector of experiential needs, EMT compels a revision of canonical economic categories such as growth, value, employment, and welfare, within a mathematically coherent, market-compatible framework. This revision calls for a reorientation of policy design, particularly where current frameworks are anchored in legacy assumptions that no longer reflect the structure of AI-enabled economic systems.

Growth-centric policy frameworks, particularly those anchored in GDP as a standalone metric, may contribute to structural lock-ins and path dependencies that perpetuate unemployment and misalignment with societal needs. Revisiting these frameworks may be essential to enable policies that reflect the broader utility landscape of AI-enabled, post-scarcity economies. EMT reinterprets growth not as the expansion of output per capita, but as the dynamic realignment of production with an expanding experiential utility frontier. The implication is a shift from quantity-optimisation to quality-alignment, implying that economic systems must evolve not toward maximum output, but toward maximal congruence between what is produced and what is meaningfully valued. EMT enables tractable policy innovation by retaining the core mechanisms of neoclassical economics, such as price-mediated coordination, rational optimisation, and decentralised decision-making, while redefining the objective function around the alignment of production with multidimensional experiential utility.

For policymakers, EMT also offers a blueprint for transitioning from GDP-based planning to alignment-based economic governance. In this paradigm, the unit of analysis is no longer a representative agent maximising utility over finite consumption bundles, but a population of agents whose experiential matrices evolve in real time, and whose needs span aesthetic, psychological, social, and existential domains. The task of economic coordination, therefore, becomes one of dynamic matching. This might be better achieved by using AI-enabled sensing and optimisation to allocate resources not by scarcity alone, but by real-time measurement of evolving human need. Certain policy measures derive from the modelling in the previous sections. 

First, one immediate policy implication concerns employment. If its theoretical assumptions hold, EMT provides formal proof that, under conditions of unmet experiential needs and idle labour, unemployment is structurally irrational, even when output is abundant. This reframes full employment not as a Keynesian demand-management policy, but as an efficiency requirement for sustaining alignment with the human utility function. Accordingly, policy should no longer treat employment purely as a cost to be minimised, but as a domain of experiential production in its own right.

To ensure humanist outcomes in employment, it is important to immediately start managing the transition to experiential alignment. While EMT models suggest that full human employment is a structural requirement in an AI-aligned, post-scarcity economy, the transition to such a configuration is unlikely to be automatic or frictionless. 

In the short term, the rise of AI-induced automation may trigger a wave of technological unemployment, not because human labour is obsolete, but because existing institutions and policy frameworks remain anchored in outdated material-output logics. EMT recasts such unemployment as a temporary structural shock, analogous in scale and urgency to the COVID-19 crisis, that requires immediate fiscal and policy intervention. 

Crucially, EMT models suggest that full human employment is not only compatible with full AI production, but is in fact a necessary feature of the experiential economy. Humans must remain embedded in economic systems to maintain alignment, ensure demand, and preserve the utility-bearing functions of agency, purpose, and participation. Thus, transition policies should be designed to stabilise employment while AI systems are reoriented toward experiential optimisation. EMT offers a formal roadmap for managing this shift, preserving decentralised coordination and market efficiency while redefining the economic objective function. Policymakers must, however, act now to steer this transition, because delaying risks deepening misalignment, institutional lock-in, and long-term utility collapse.

Second, EMT justifies the redirection of fiscal and industrial policy toward AI systems that enhance ideation, coordination, and affective mapping. These are not peripheral functions, but the general-purpose capabilities through which the economy can traverse new experiential dimensions. Rather than subsidising output in traditional sectors, governments should invest in alignment infrastructure, such as platforms, protocols, and institutions capable of translating high-dimensional human feedback into economic action. This includes the design of ethical AI architectures, data governance frameworks, and participatory sensing mechanisms that ensure production systems remain accountable to the experiential realities of those they serve.

Third, EMT invites a wholesale reevaluation of welfare economics. Traditional measures such as GDP per capita or utility from consumption for its own sake are inadequate in a world where utility is decoupled from material scarcity. EMT’s convergence results imply that welfare should be measured not by what is consumed, but by the degree of alignment between economic systems and the full experiential matrix \(E(t)\). This suggests a new class of alignment-based welfare metrics, including for example experiential unmet need indices, satisfaction-convergence indicators, and misalignment gap diagnostics, each of which could be derived from the formal architecture of EMT and operationalised through AI-enabled sensing.

Finally, EMT offers a theoretical justification for anticipatory economic policy. Because ideation costs decline endogenously with AI development, policy lags pose a structural risk of misalignment. Waiting for market failures to manifest before intervening allows irreversible gaps in utility to emerge. EMT therefore legitimises proactive governance, and the potential for aligning regulation, investment, and institutional design with the trajectory of AI-induced transformation. This includes strategic forecasting of emergent needs, simulation of post-scarcity equilibria, and the coordination of public-private transitions to meaning-centred production regimes.

In sum, EMT transforms economics from a logic of equilibrium allocation to one of dynamic alignment. It preserves the fundamental tools of economics, such as optimisation, decentralised exchange, and price adjustment, but rewires the architecture around a new object of maximisation, the evolving, irreducible matrix of human experience. In doing so, it offers not merely a critique of existing paradigms, but a mathematically grounded and policy-relevant theory for designing economies that flourish as AI advances.

\subsection{Strategic Roadmap for Future Research in Alignment Economics}

The introduction of EMT opens an expansive research frontier we term alignment economics, a potential new field dedicated to studying economic systems that continuously align output with the full spectrum of human experiential needs, particularly in the context of accelerating AI capabilities. Given the importance of transcending the paradigm of GDP-oriented growth and to offer a coherent alternative, this emerging field calls for an academic platform that integrates formal economic methods with ethical reasoning, institutional theory, and AI systems design. It also demands a coordinated programme of theoretical, empirical, and applied research. We now outline five core research domains that may be necessary for advancing this agenda.

First, operationalising the experiential matrix may need to become a core focus of research and policy. A foundational task is the empirical and computational characterisation of the experiential matrix \( E(t) \). This requires interdisciplinary collaboration to identify, define, and weight experiential dimensions such as purpose, belonging, coherence, and mental health. Techniques from affective computing, psychometrics, and behavioural science must be integrated into economic modelling to map satisfaction states \( x_i(t) \) with rigor. Machine learning models trained on multimodal data (e.g., biometric, linguistic, behavioural) can assist in constructing real-time proxies for utility vectors. Research in this stream should focus on formalising experience-based utility functions and validating them against observable indicators of flourishing.

Second, AI-aligned control systems need to be used to derive tractable approximations to guide EMT alignment. The infinite-dimensional optimal control structure presented in EMT invites computational development. Future work should derive finite, tractable approximations of the EMT Hamiltonian, allowing simulation and calibration of aligned production systems. This includes extending existing dynamic stochastic general equilibrium (DSGE) models to incorporate ideation cost decay, AI-driven production mappings, and endogenous feedback between satisfaction and innovation. Applications may range from national planning models to firm-level decision-making architectures. This line of research can benefit from advances in high-dimensional control theory, reinforcement learning, and AI planning under uncertainty.

Third, alignment metrics should be developed together with policy instruments. Building on EMT’s convergence theorems, a new class of economic metrics is needed to evaluate alignment quality and utility expansion over time. Future research should explore how such metrics can inform dynamic policy tools such as tax incentives, public investment, or algorithmic governance, that maintain alignment and pre-empt systemic misalignment. The design of such policy instruments will require interdisciplinary methods integrating welfare economics, behavioural insights, and ethical AI principles.

Fourth, institutional and organisational design thinking needs to be updated for a potential era of post-scarcity economies. EMT implies that economic coordination must evolve from efficiency-maximising structures to alignment-sustaining architectures. This opens new lines of inquiry in institutional economics, organisational theory, and political economy. Future research should examine how existing institutions, e.g., central banks, regulatory bodies, and education systems, can be repurposed to act as \textit{alignment mediators}. Future suggested questions include: What governance models best ensure participatory sensing of needs? How can markets internalise alignment externalities? What is the role of AI cooperatives, commons-based infrastructures, or citizen feedback protocols in dynamically updating utility functions? This area also demands novel theory-building on incentive compatibility and legitimacy in AI-led coordination systems.

The fifth core research domain enabling the EMT alignment transition requires revisiting certain philosophical foundations and notions of normative pluralism. The normative core of EMT, which implies that utility must include meaning, coherence, and self-authorship, calls for engagement with moral philosophy, value theory, and epistemology. Future work must explore how different societies prioritise experiential dimensions, how weights \( w_i \) might be contextually constructed, and how plural utility architectures can coexist without collapsing into relativism or incoherence. This includes research on deliberative valuation methods, culturally-sensitive AI design, and normative robustness in welfare aggregation. EMT thereby offers a formal structure within which ethical diversity can be modelled rather than abstracted away.

\medskip

Together, these five domains might constitute the foundational research programme of alignment economics. As AI accelerates the collapse of ideation and coordination costs, the imperative is not merely to react, but to design. EMT provides the analytical scaffolding. Future research must now construct the models, metrics, and mechanisms through which societies can navigate the post-scarcity transition. If successful, this field may do for the AI era what Keynesianism did for the industrial age, to reorient economic purpose to serve the full flourishing of humanity.

\subsection{Concluding Reflection on EMT and Human Flourishing}

EMT is not merely a reformulation of utility or a technical proposal for aligning production systems, but a theoretical lens through which the future of economic life may be reimagined. In a world increasingly shaped by AI, traditional notions of value, growth, and employment begin to lose coherence unless re-anchored in the evolving structure of human experience. EMT provides that anchor, offering a framework in which economic reasoning, technological progress, and moral purpose can converge.

What this paper has suggested is that utility, when properly conceived from a human or humanist vantage point, is not static, finite, or reducible to consumption. It is experiential, irreducible, and infinite in dimensionality. This reconceptualisation is not speculative but mathematically rigorous, economically tractable, and philosophically grounded. By proving, based on EMT's assumptions, that AI can be deployed to collapse ideation costs and systematically close the gap between production and lived human experience, EMT reframes AI not as an external disruptor, but as a structural enabler of meaning, coherence, and flourishing.

Importantly, EMT does not reject markets, price systems, or decentralised coordination. It affirms their foundational roles while reprogramming their objective function. In this view, the economy is no longer a machine for allocating scarcity but a sensorimotor system for navigating the infinite topology of human need. The objective of economic design, therefore, is not maximum output, but maximum alignment, between what is technologically possible and what is humanly meaningful.

This shift has existential consequences. As AI approaches general-purpose capability, societies must choose what it is for. EMT provides a formal answer, that AI is for the alignment of economic systems with the full structure of human utility, including its most abstract, intangible, and transcendent dimensions. It is for freedom, not just from want, but toward meaning. It is for securing not only material well-being, but ontological continuity in a time of accelerating change.

If the theorems of EMT hold, and they are constructed to withstand rigorous scrutiny, then a new phase of economics becomes possible. One in which unemployment is no longer rational, because misalignment is no longer tolerated. One in which growth is redefined as the progressive resolution of unmet experiential needs. One in which the economy itself becomes an instrument for human development in its most expansive sense.

This paper has laid a theoretical groundwork for the alignment of humanist experiential needs with productive forces. The next task is collaborative. Researchers, policymakers, technologists, and citizens must now build the institutional, ethical, and computational infrastructures required to realise this vision. The challenge is substantial, but so is the opportunity. EMT suggests that the future of economic life need not be one of redundancy, alienation, or inequality. It can instead be one of purpose, participation, and shared flourishing, if we choose to align the systems we build with the lives we seek to live.

\newpage
\noindent \textbf{References}

Acemoglu, D. and P. Restrepo (2019). Artificial intelligence, automation, and work. The economics of artificial intelligence: An agenda. pp. 197-236. Chicago: University of Chicago Press.

Arrow, K. J. and G. Debreu (1954). 'Existence of an equilibrium for a competitive economy', Econometrica (pre-1986), 22, pp. 265-290.

Arthur, W. B. (1989). 'Competing technologies, increasing returns, and lock-in by historical events', The Economic Journal, 99, pp. 116-131.

Balder, E. J. (1983). 'An existence result for optimal economic growth problems', Journal of Mathematical Analysis and Applications, 95, pp. 195-213.

Becker, G. S. (1976). The economic approach to human behavior, University of Chicago press, Chicago.

Bresnahan, T. F. and M. Trajtenberg (1995). 'General purpose technologies ‘Engines of growth’?', Journal of Econometrics, 65, pp. 83-108.

Cass, D. (1965). 'Optimum Growth in an Aggregative Model of Capital Accumulation1', The Review of Economic Studies, 32, pp. 233-240.

Debreu, G. (1954). Representation of a preference ordering by a numerical function. Decision processes. pp. 159-165. New York: John Wiley and Sons.

Debreu, G. (1959). Theory of value: An axiomatic analysis of economic equilibrium, Yale University Press, New Haven.

Frankl, V. E. (1985). Man's search for meaning, Simon and Schuster, New York.

Galbraith, J. K. (1998). The Affluent Society, Houghton Mifflin Harcourt, Boston.

Grossman, G. M. and E. Helpman (1991). 'Quality Ladders in the Theory of Growth', The Review of Economic Studies, 58, pp. 43-61.

Jones, C. I. (1995). 'R\&D-based models of economic growth', Journal of Political Economy, 103, p. 759.

Koopmans, T. C. (1965). On the concept of optimal economic growth. The Economic Approach to Development Planning. Amsterdam: North-Holland Publishing Co.

Nussbaum, M. C. (2007). Capabilities as fundamental entitlements: Sen and social justice. Capabilities equality. pp. 54-80. Milton Park: Routledge.

Pontryagin, L. S., V. G. Boltyanskii, R. V. Gamkrelidze and E. F. Mishchenko (1962). The Mathematical Theory of Optimal Processes, Inderscience, New York.

Ramsey, F. P. (1928). 'A Mathematical Theory of Saving', The Economic Journal, 38, pp. 543-559.

Romer, P. M. (1990). 'Endogenous technological change', Journal of Political Economy, 98, pp. S71-S102.

Samuelson, P. (1947). Foundations of Economic Analysis, Harvard University Press, Cambridge.

Scitovsky, T. (1976). The joyless economy: An inquiry into human satisfaction and consumer dissatisfaction, Oxford University Press, Oxford, England.

Sen, A. (1996). 'Freedom favors development', New Perspectives Quarterly, 13, pp. 23-27.

von Neumann, J., O. Morgenstern and A. Rubinstein (1944). Theory of Games and Economic Behavior, Princeton, Princeton University Press.

\end{document}